\documentclass[journal=jacsat,manuscript=article]{achemso}

\usepackage[version=3]{mhchem} 
\usepackage[T1]{fontenc}       
\usepackage{amsmath}
\usepackage{subfig}
\usepackage{graphics}
\usepackage{epstopdf}

\author{Shiyang Long}
\affiliation[School of Life Sciences, Jilin University, Changchun, China 130012]
{School of Life Sciences, Jilin University, Changchun, China 130012}
\author{Pu Tian}
\email{tianpu@jlu.edu.cn}
\phone{+86 (0)431 85155287}
\affiliation[School of Life Sciences, Jilin University, Changchun, China 130012]
{School of Life Sciences, Jilin University, Changchun, China 130012}
\alsoaffiliation[MOE Key Laboratory of Molecular Enzymology and Engineering, Jilin University, Changchun, China 130012]
{MOE Key Laboratory of Molecular Enzymology and Engineering, Jilin University, Changchun, China 130012}

\title[]{Nonlinear backbone torsional pair correlations in proteins}

\abbreviations{}

\begin{document}

%
%
%
%
%

\begin{abstract}
Protein allostery requires dynamical structural correlations. Physical origin of which, however, remain elusive despite intensive studies during last two decades. Based on analysis of molecular dynamics (MD) simulation trajectories for ten proteins with different sizes and folds, we found that nonlinear backbone torsional pair (BTP) correlations, which are spatially more long-ranged and are mainly executed by loop residues, exist extensively in most analyzed proteins. Examination of torsional motion for correlated BTPs suggested that aharmonic torsional state transitions are essential for such non-linear correlations, which correspondingly occur on widely different and relatively longer time scales. In contrast, BTP correlations between backbone torsions in stable $\alpha$ helices and $\beta$ strands are mainly linear and spatially more short-ranged, and are more likely to associate with intra-well torsional dynamics. Further analysis revealed that the direct cause of non-linear contributions are heterogeneous, and in extreme cases canceling, linear correlations associated with different torsional states of participating torsions. Therefore, torsional state transitions of participating torsions for a correlated BTP are only necessary but not sufficient condition for significant non-linear contributions. These findings implicate a general search strategy for novel allosteric modulation of protein activities. Meanwhile, it was suggested that ensemble averaged correlation calculation and static contact network analysis, while insightful, are not sufficient to elucidate mechanisms underlying allosteric signal transmission in general, dynamical and time scale resolved analysis are essential. 
\end{abstract}

\section{Introduction}
\begin{figure}
\centering 
\subfloat[1bta]{\includegraphics[width=1.2in,angle=90]{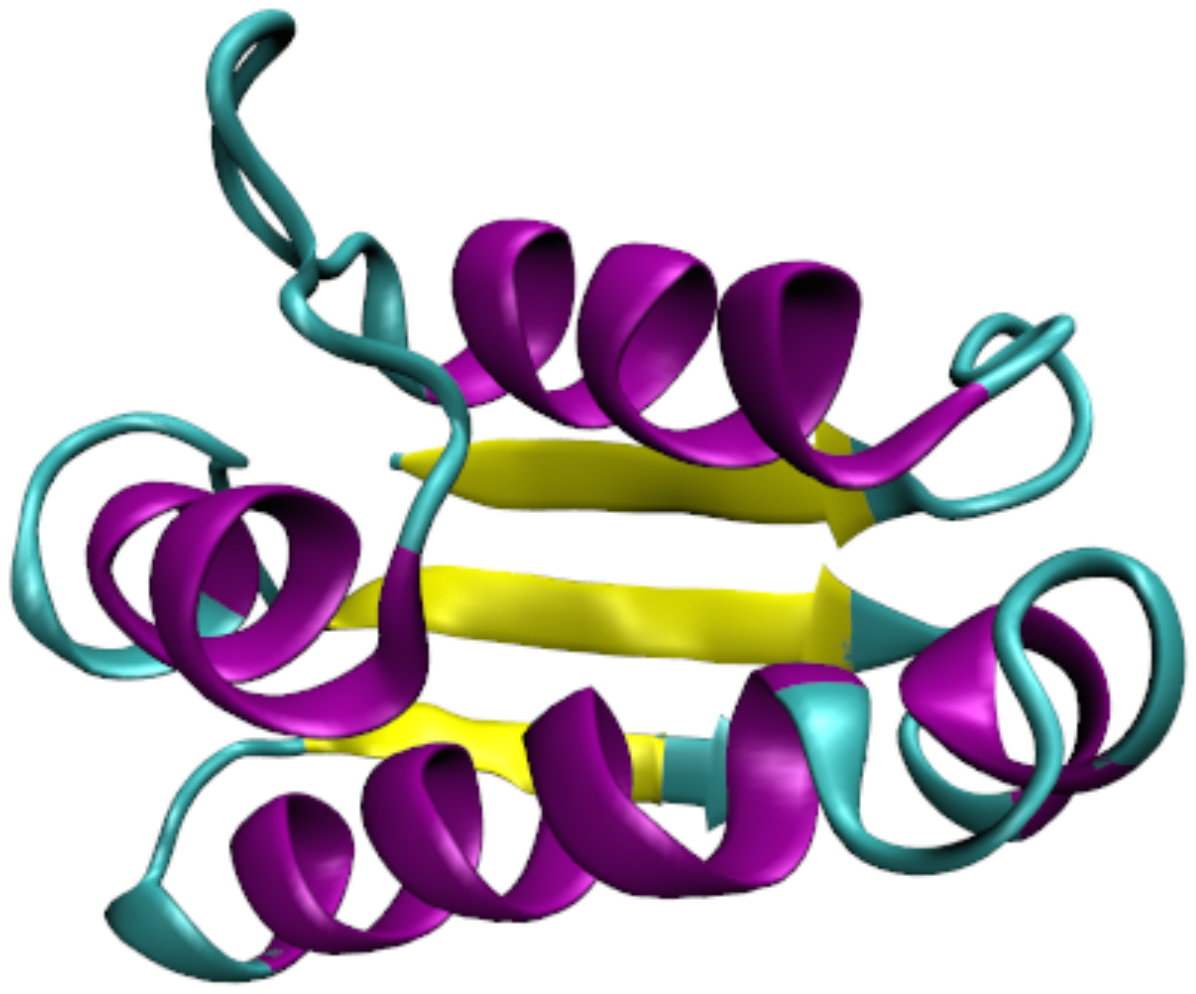}}
\subfloat[1rgh]{\includegraphics[width=1.2in]{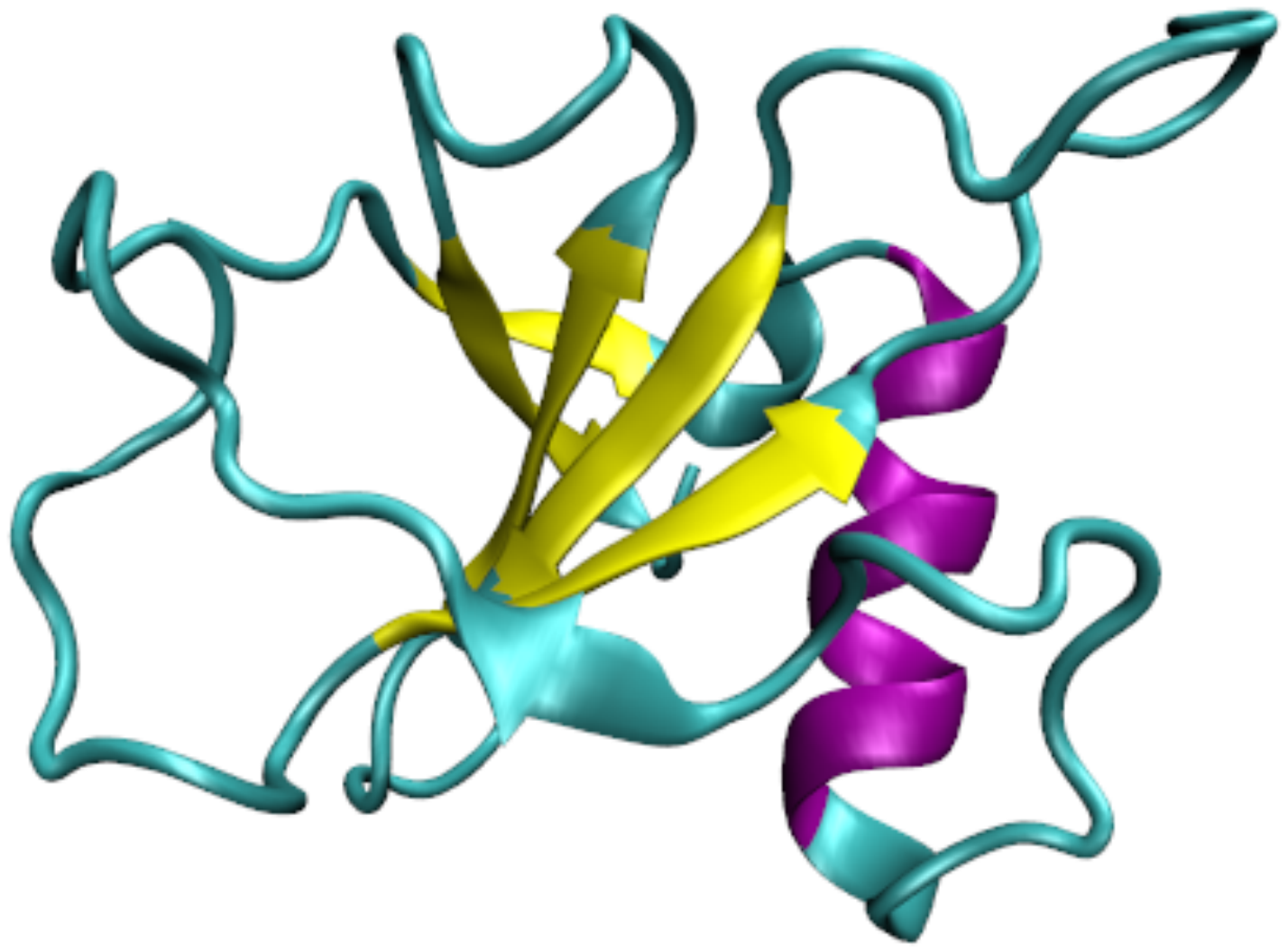}}
\subfloat[2bnh]{\includegraphics[width=1.2in]{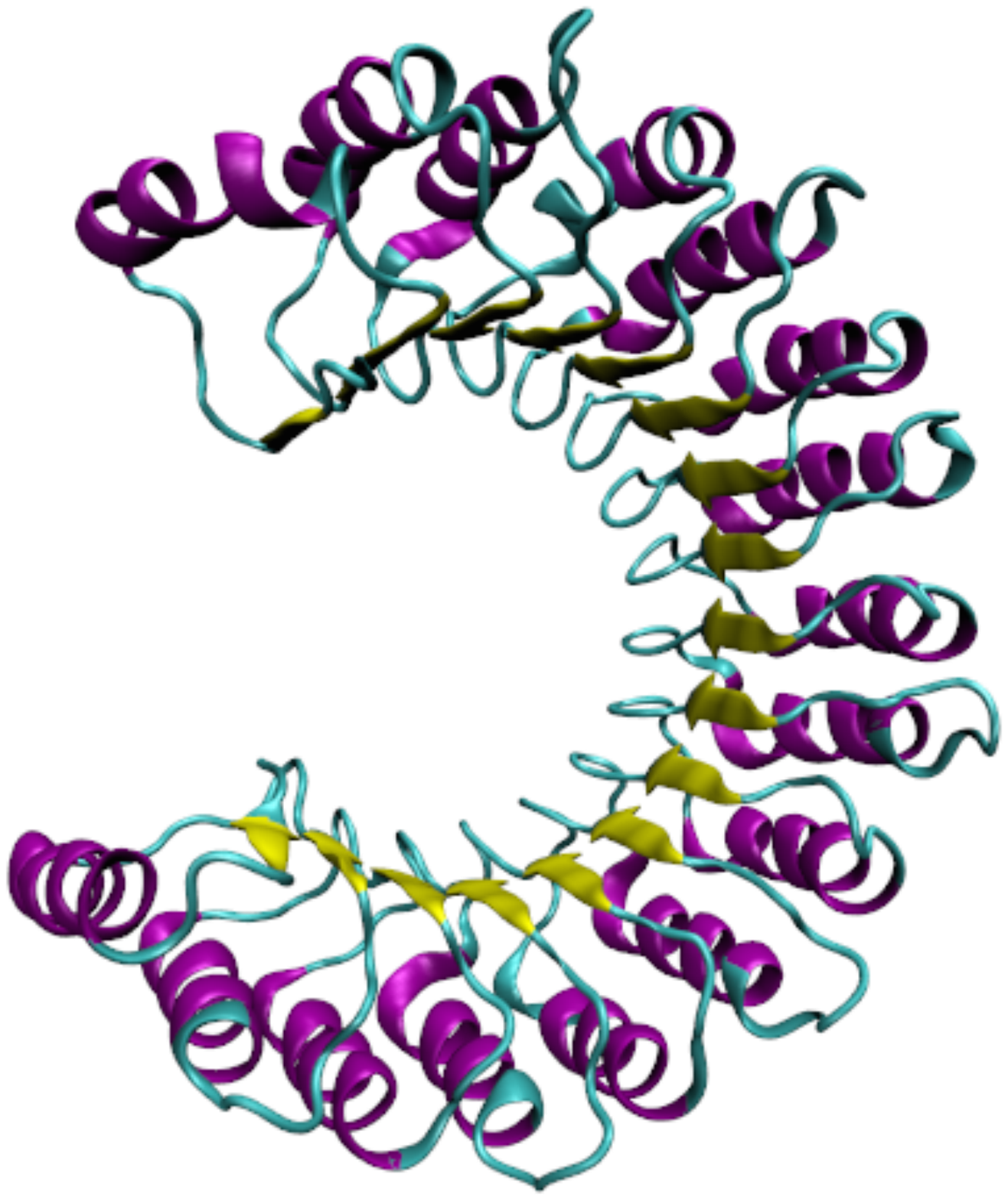}}
\subfloat[2pka]{\includegraphics[width=1.2in]{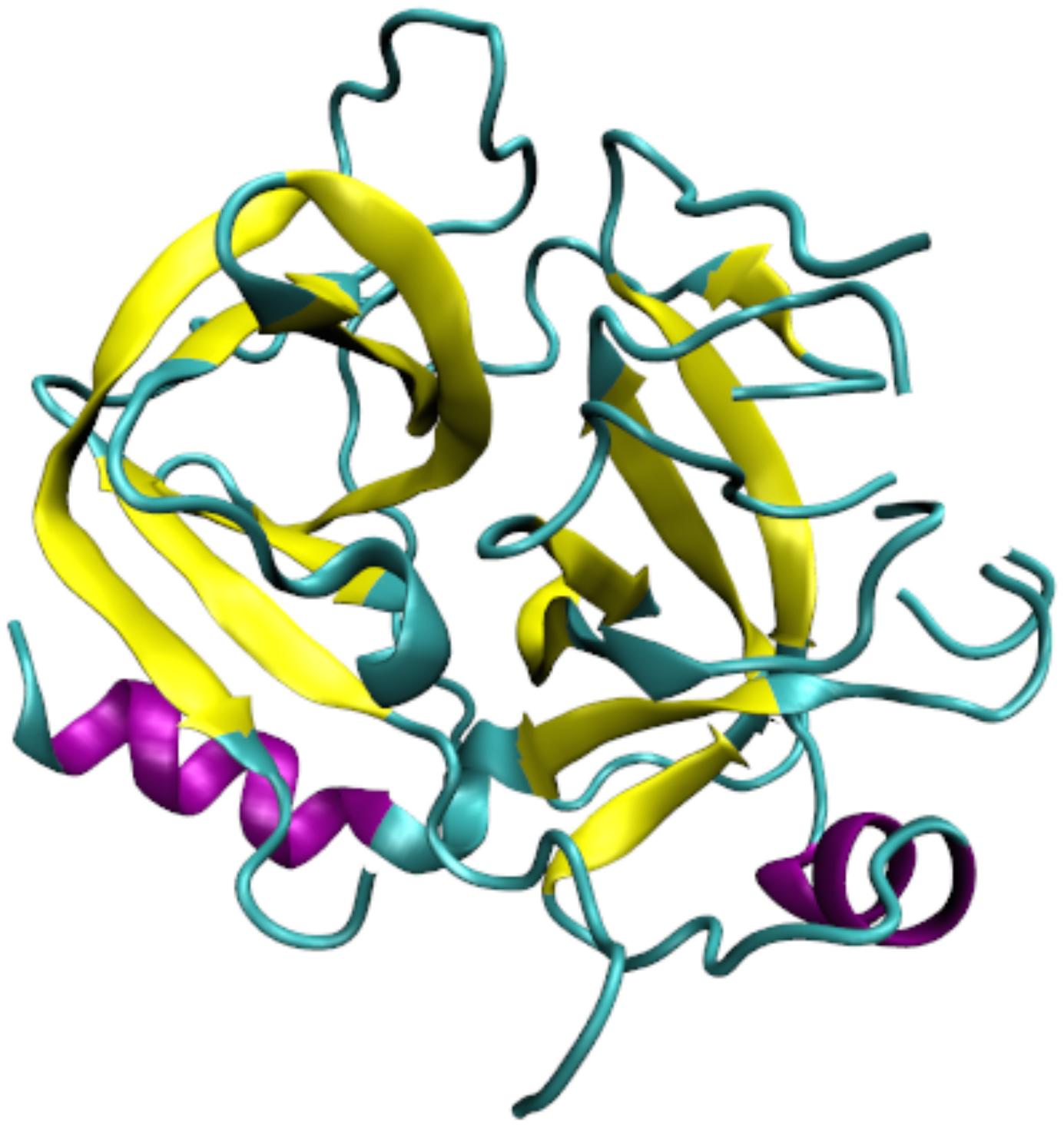}}
\subfloat[3f3y]{\includegraphics[width=1.2in]{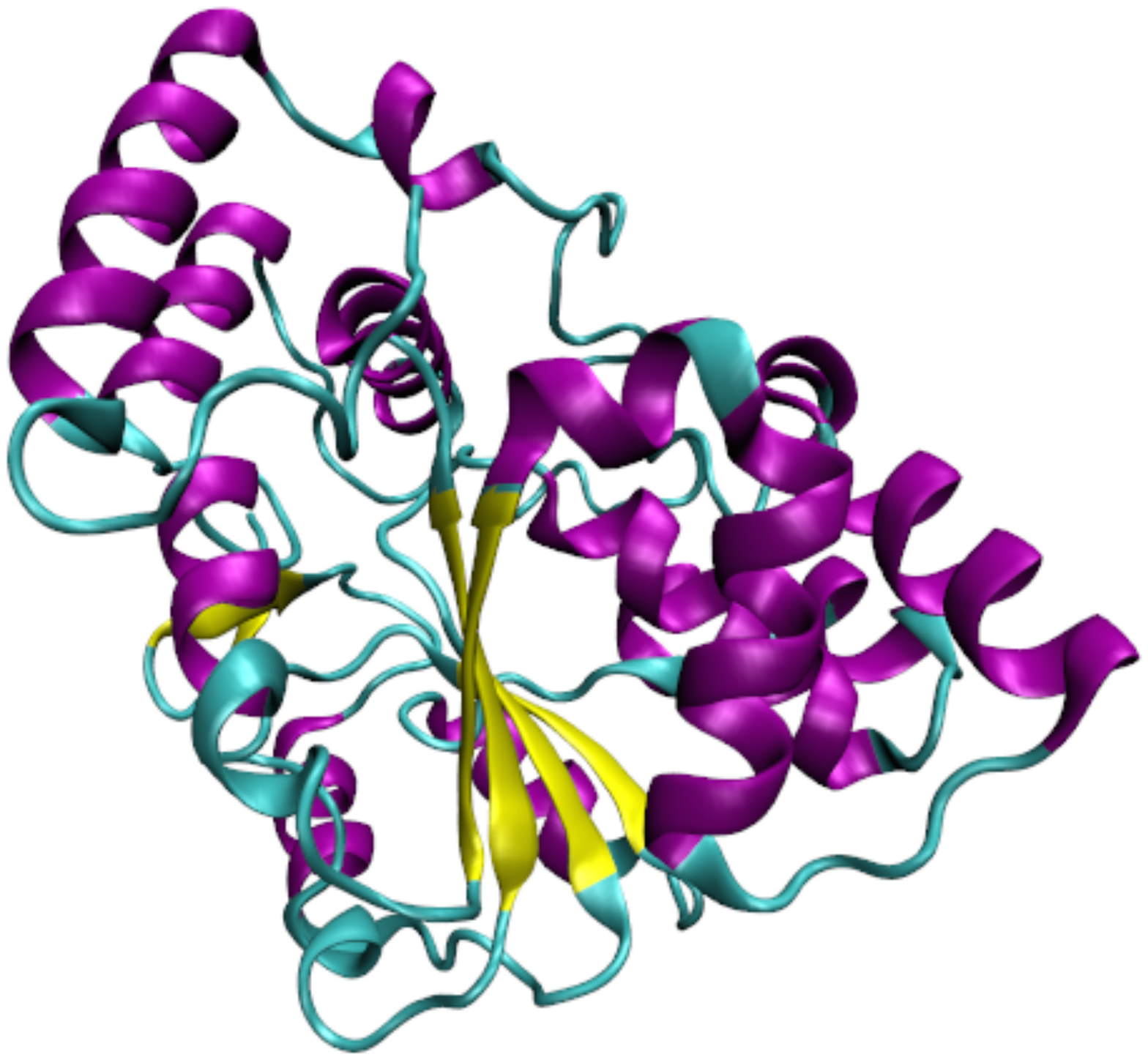}}\\
\subfloat[5pti]{\includegraphics[width=1.2in]{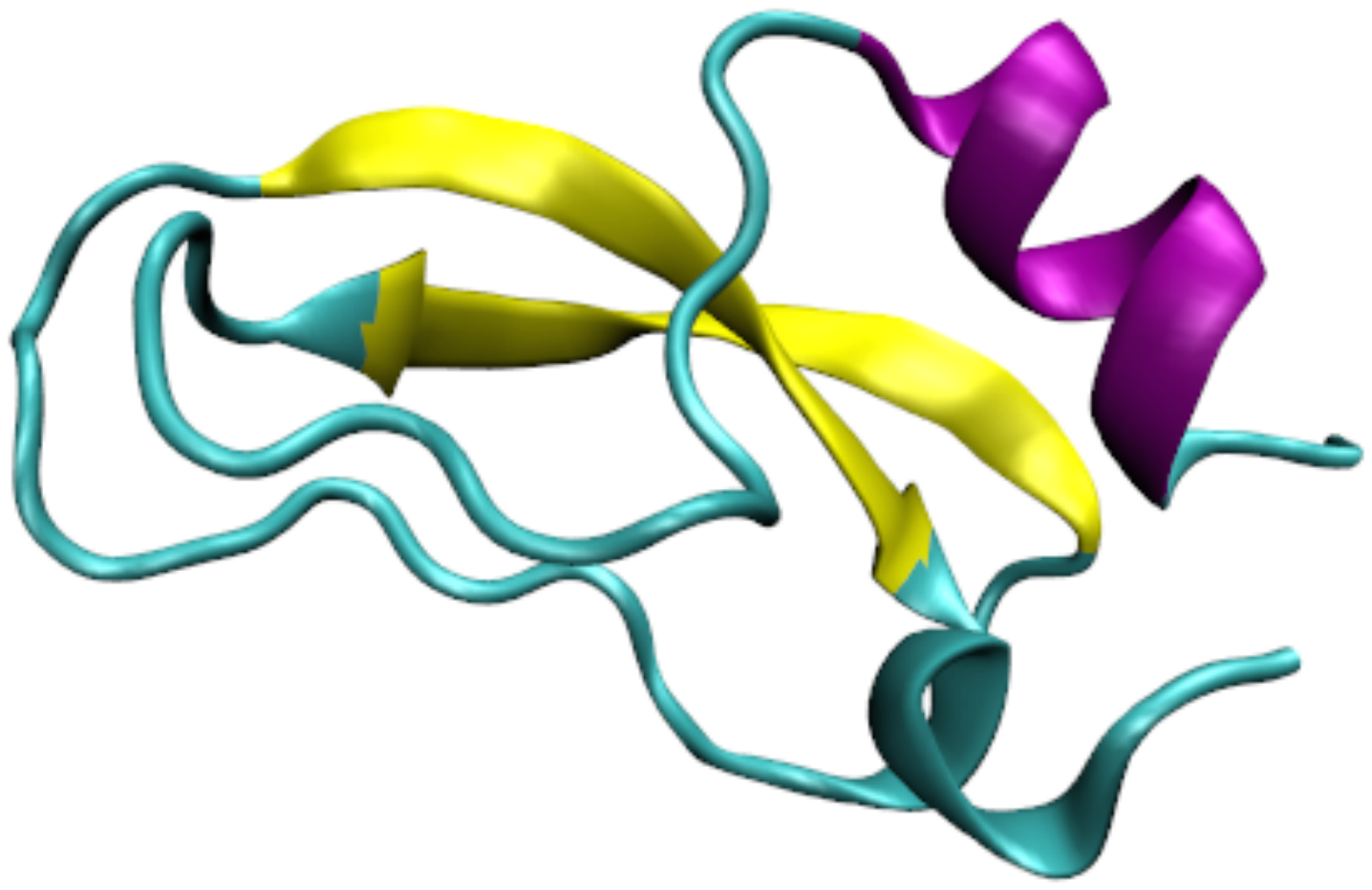}}
\subfloat[7rsa]{\includegraphics[width=1.2in]{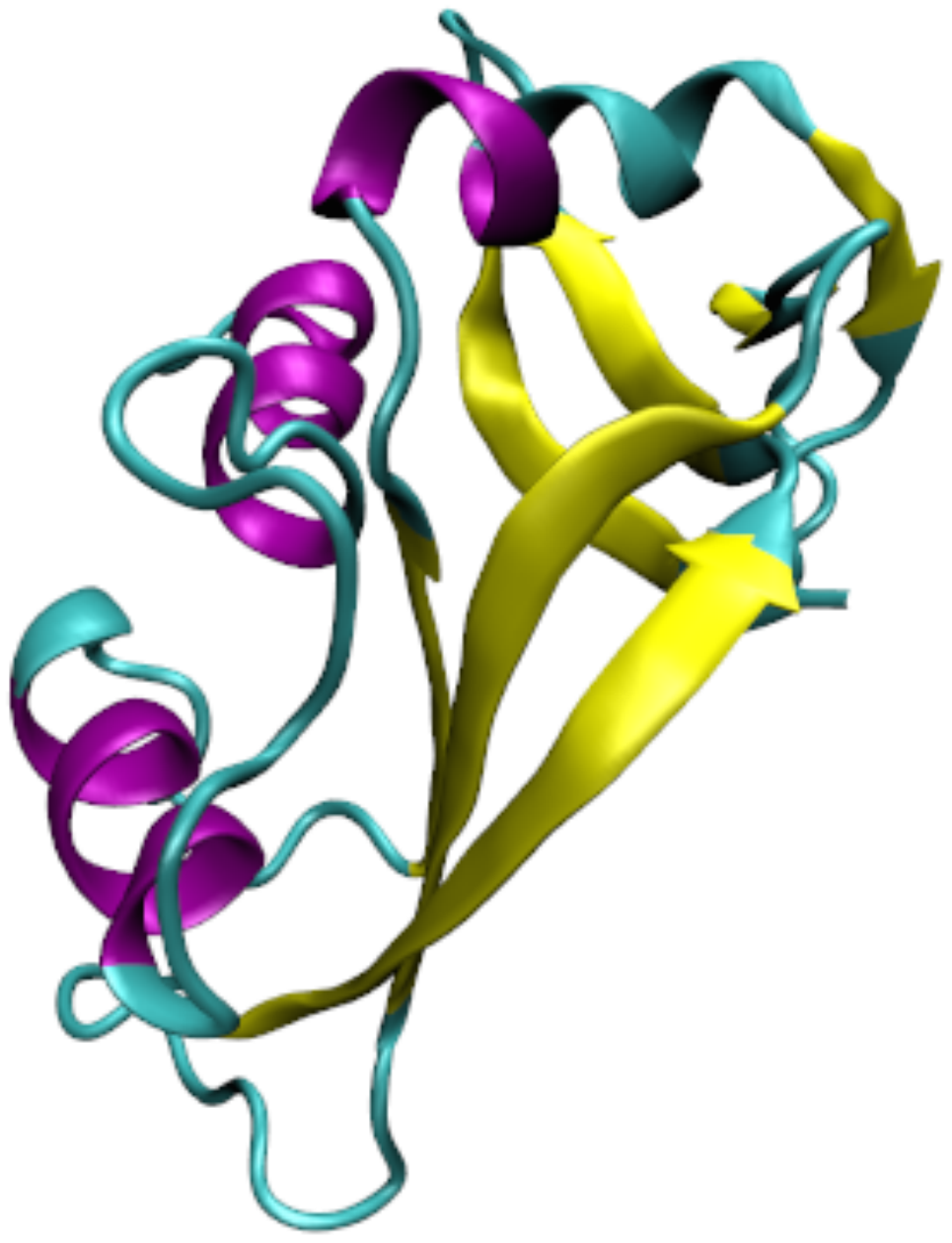}}
\subfloat[Bame]{\includegraphics[width=1.2in]{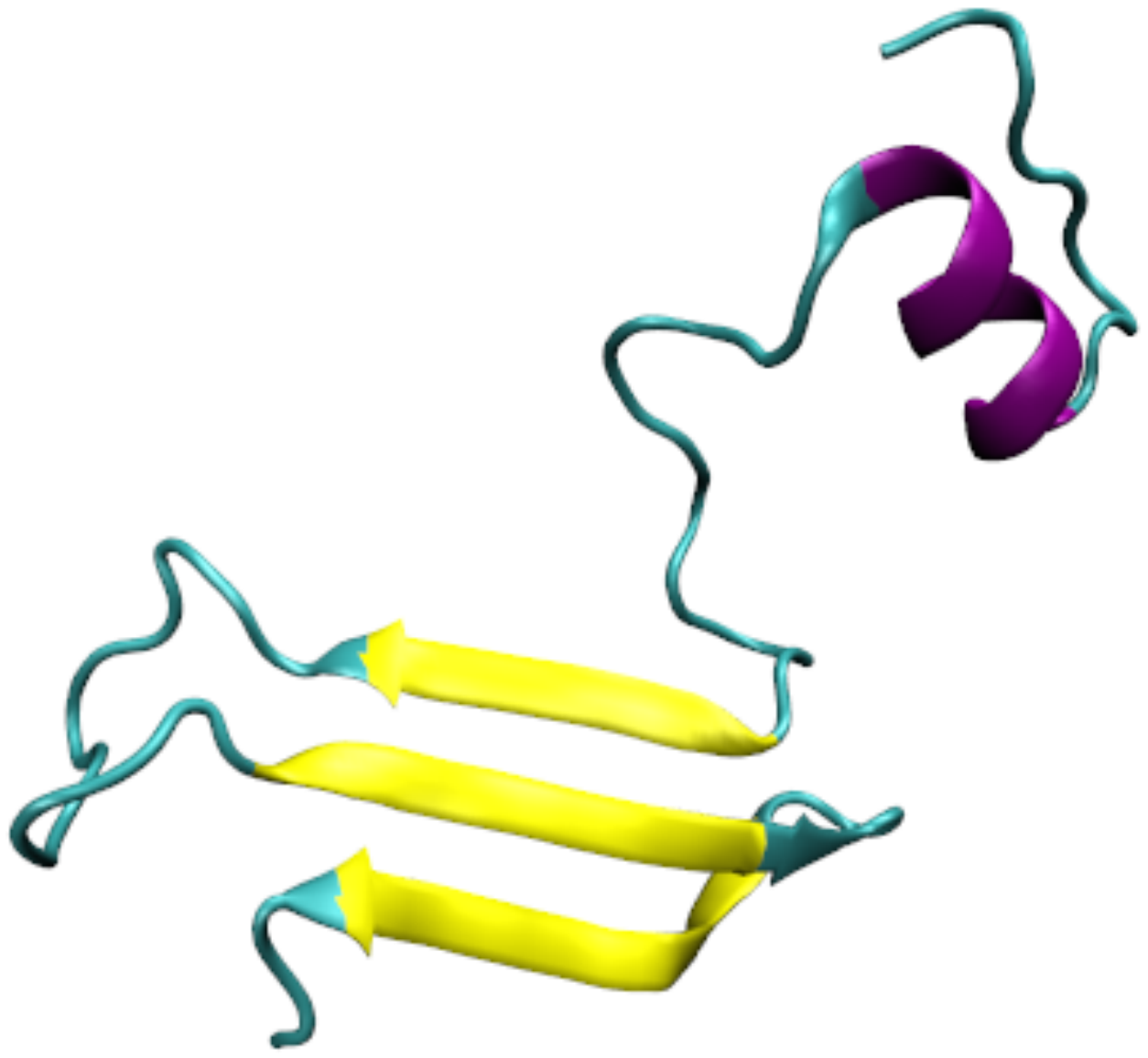}}
\subfloat[CDK2]{\includegraphics[width=1.2in]{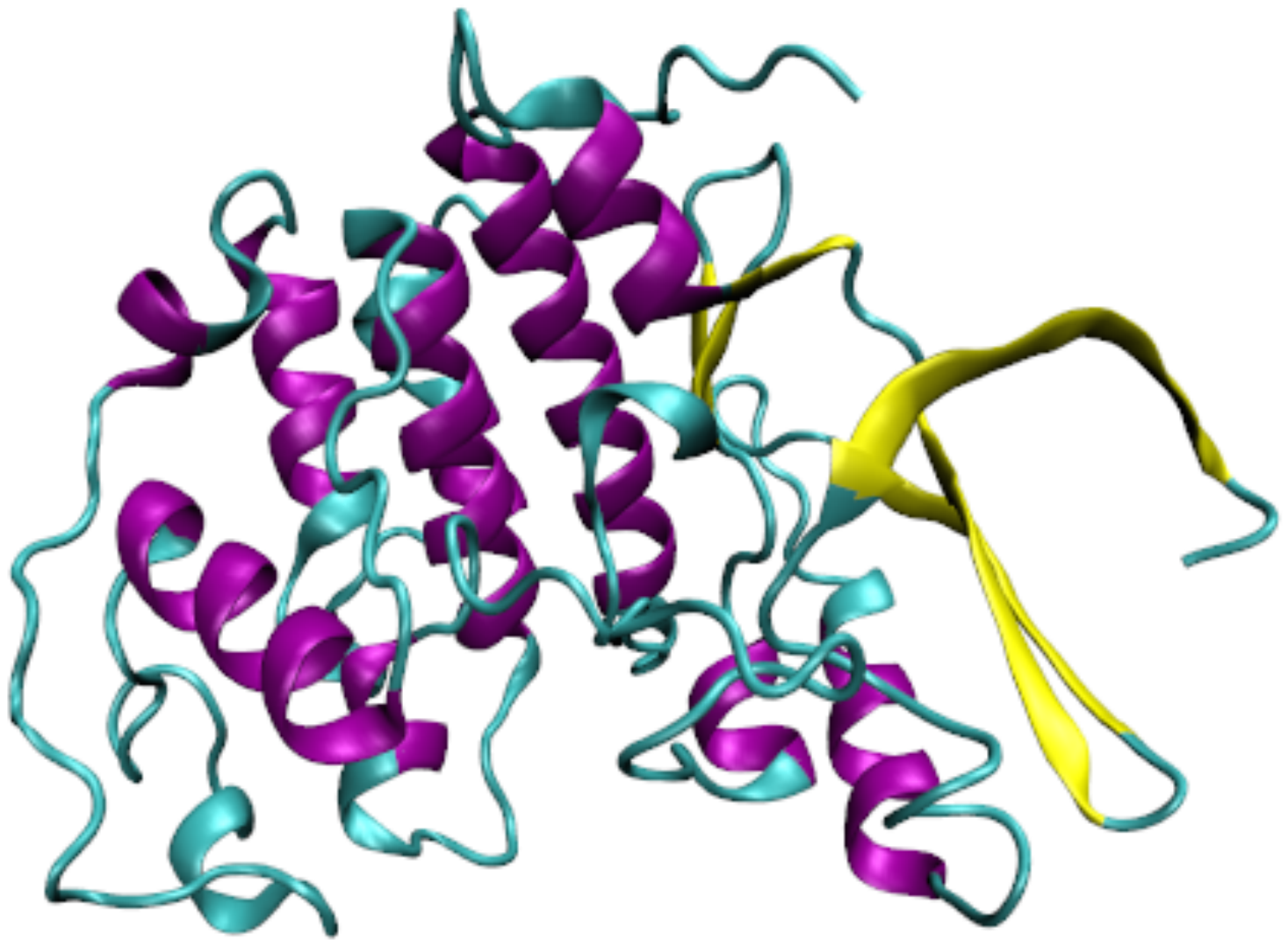}}
\subfloat[lyzm]{\includegraphics[width=1.2in]{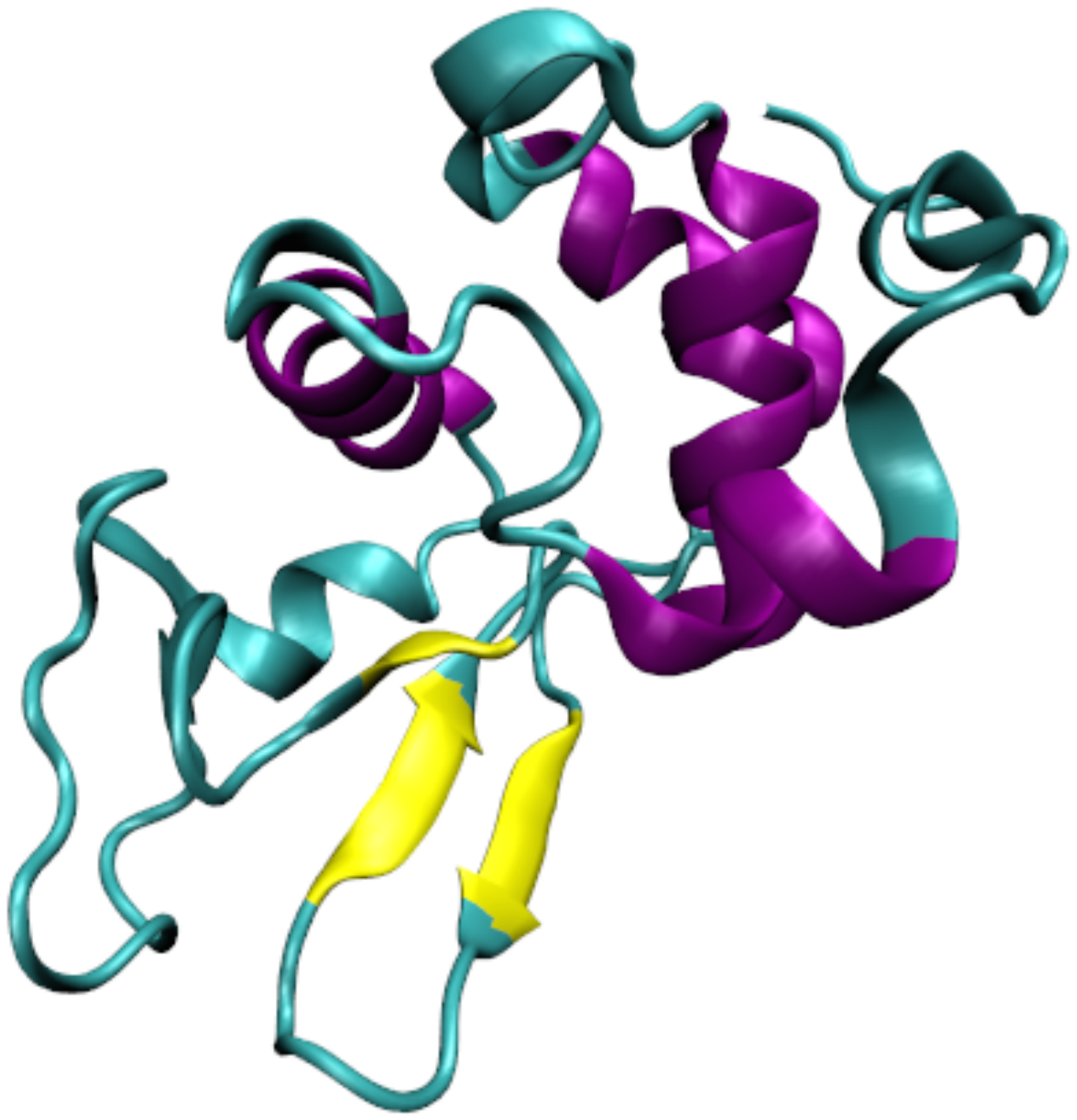}}
\caption{Structures of ten proteins (labeled with PDB codes) analyzed in this study, $\alpha$ helices are in purple, $\beta$ strands are in yellow, all other secondary structures were termed ``loop'' in this study and are shown in cyan. Figures were prepared with VMD\cite{VMD}} 
\label{fig:structure}
\end{figure}

Allostery in protein molecules is defined by their response to external stimuli on distal site(s). Most biologically relevant allostery are spatially long-ranged\cite{Brunori2011,Roy2012,Nussinov2013,Nussinov2013a,Motlagh2014}. Therefore, understanding of structural correlations, especially long-ranged ones, are essential for elucidation and manipulation of protein allostery. Earlier computational characterization of dynamical correlations\cite{Ichiye1991, Hunenberger1995, Garnier1996}, despite limited time scales, provide insightful information on the inherent correlated motion and response of a transmembrane helical peptide to an exerted local conformational change. Li \emph{et. al.}\cite{Li2009a} analyzed a 700-$ns$ molecular dynamics (MD) simulation trajectory of ubiquitin 
and concluded that long-ranged pair correlations are rather rare and network of short-ranged coherent motions likely contribute to transmission of information in allostery.  By combining NMR and computational ensemble, Fenwick \emph{et. al.} \cite{Fenwick2011} concluded that observed limited long-range correlations in ubiquitin are likely to be transmitted by network of hydrogen bonds. Along the same line, Fenwick \emph{et. al.}\cite{Fenwick2014} provided evidence that hydrogen bonds across $\beta$-sheets mediates concerted motions, which are candidates for transfer of structural information over relatively long distances. Papaleo \emph{et. al.} combined a description of the protein as a network of interacting residues and dynamical cross-correlation to detect communication pathways from MD simulation trajectories of E2 enzymes\cite{Papaleo2012}. In these studies, analyses were limited to linear correlations\cite{Ichiye1991, Hunenberger1995, Garnier1996,Li2009a,Fenwick2011,Papaleo2012,Fenwick2014}. It was well-recognized that non-linear correlations exist in protein dynamics and a generalized correlation measure was developed to be within the range of [0,1] based on non-linear transformation of mutual information (MI)\cite{Lange2006}. A procedure of mutual information based correlation analysis was developed and utilized to identify long-range correlations in human interleukin-2.
 However, despite important insight revealed in these studies, the physical origin and underlying molecular motions of observed correlations remain elusive. In this study, we focus on molecular motions that underly backbone torsional pair (BTP) correlations. After calculating both mutual information and linear correlations for BTPs in extensive MD simulation trajectories of ten proteins with different sizes and folds (Fig. \ref{fig:structure}), we analyzed variation of correlations as a function of sequential and spatial distances, and as a function of belonging secondary structures, torsional motions and time scales. It was found that linear correlations of BTPs are predominantly spatially short-ranged, mainly associate with harmonic local torsional motions and occur on relatively short time scales. On the other hand, non-linear correlations occur for both spatially short and long-ranged BTPs, they associate \emph{exclusively} with aharmonic torsional state transitions on widely different and relatively longer time scales, and are dominantly executed by loop residues. However, for a correlated BTP, torsional state transitions are necessary but not sufficient for non-linear correlations. The direct cause of non-linear BTP correlations are found to be heterogeneous, or in extreme cases canceling, linear correlations associated with different torsional states of participating torsions.

\section{Results}
\subsection{Mutual information and linear correlations of BTPs}
\begin{figure}
\centering 
\subfloat[1rgh]{\includegraphics[width=3.5in]{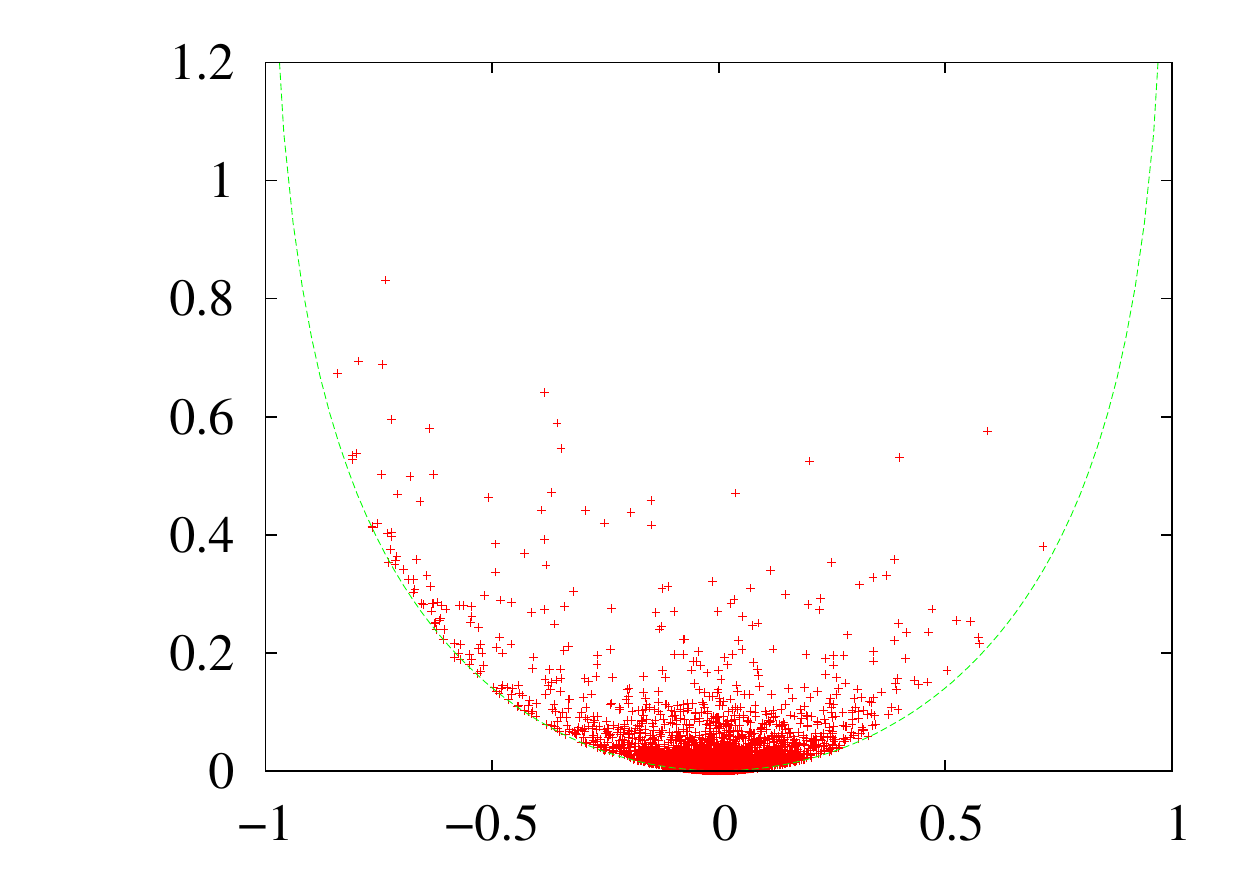}}
\subfloat[7rsa]{\includegraphics[width=3.5in]{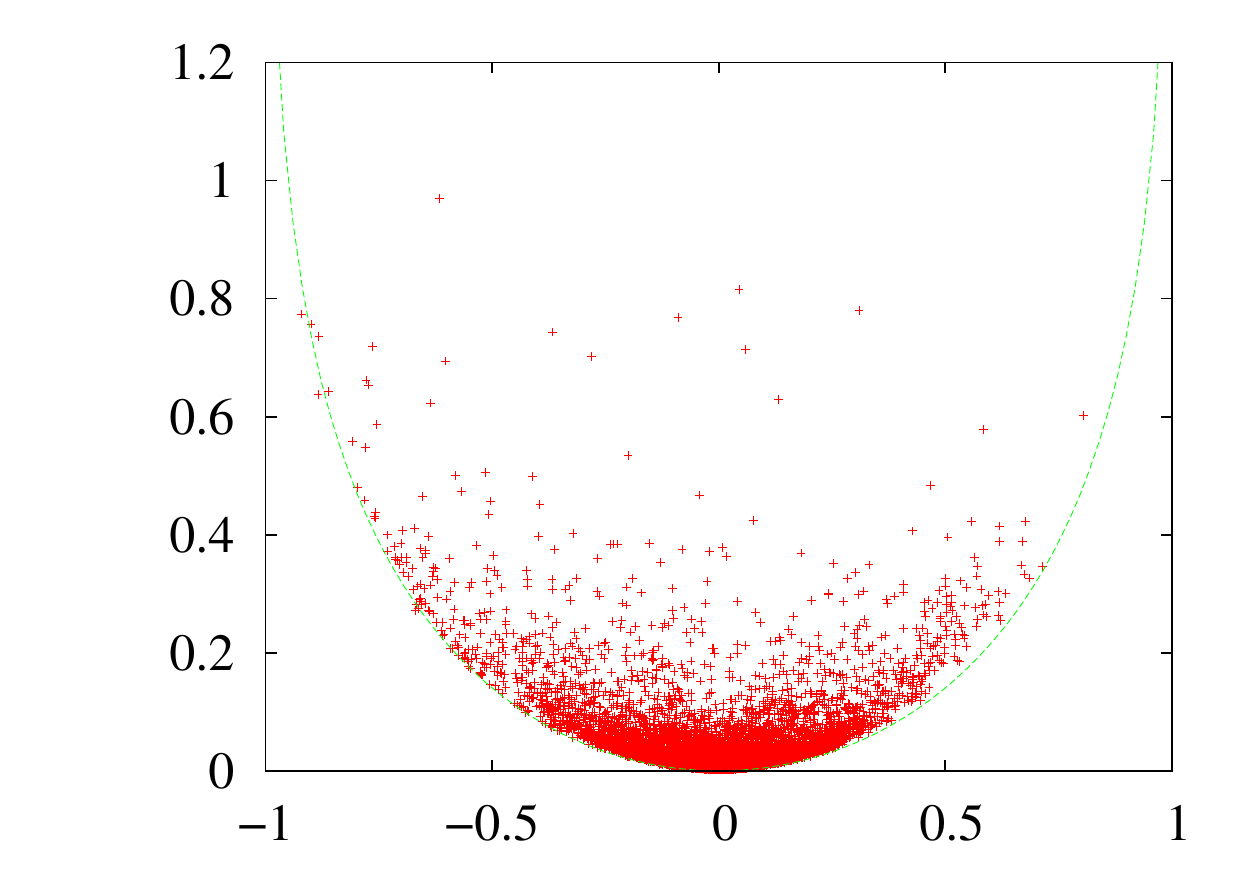}}\\
\subfloat[cdk2]{\includegraphics[width=3.5in]{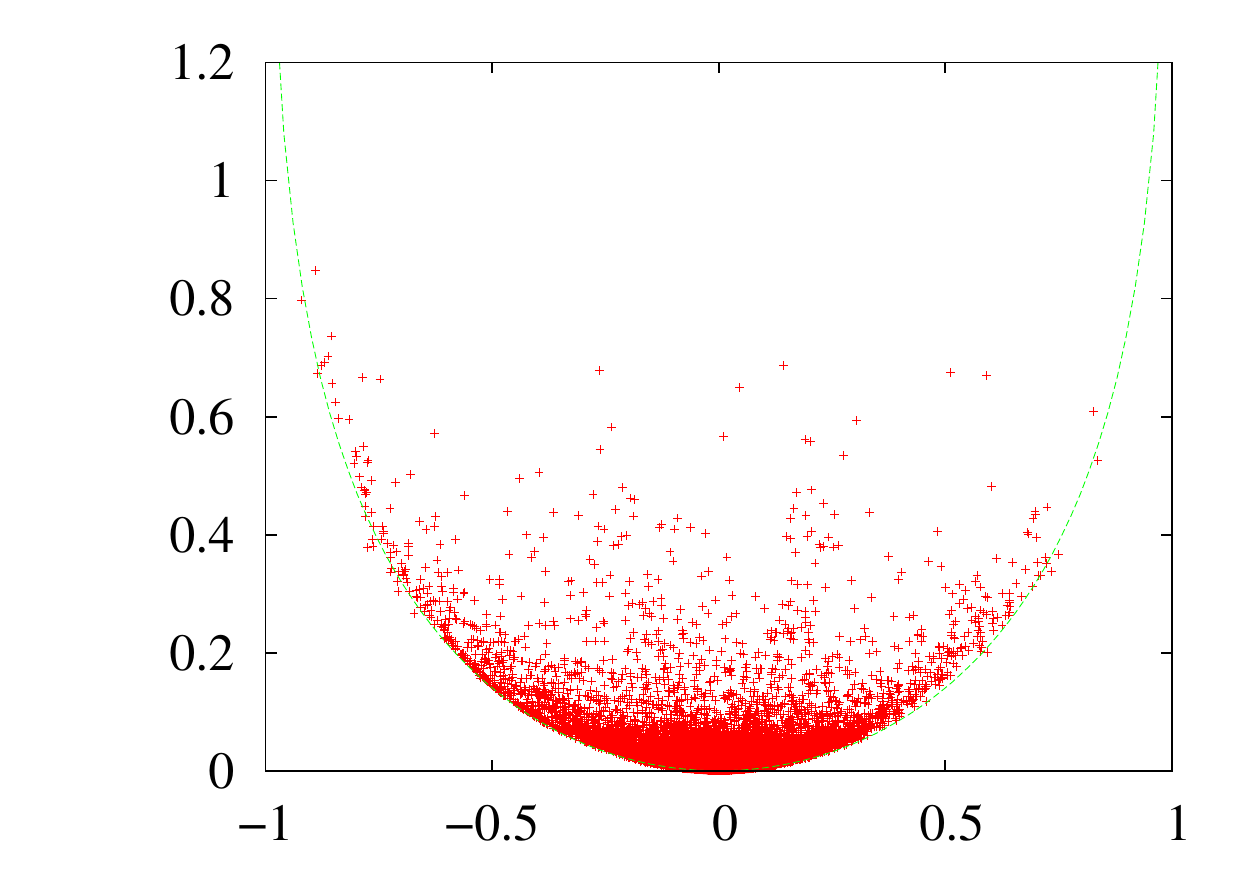}}
\subfloat[lyzm]{\includegraphics[width=3.5in]{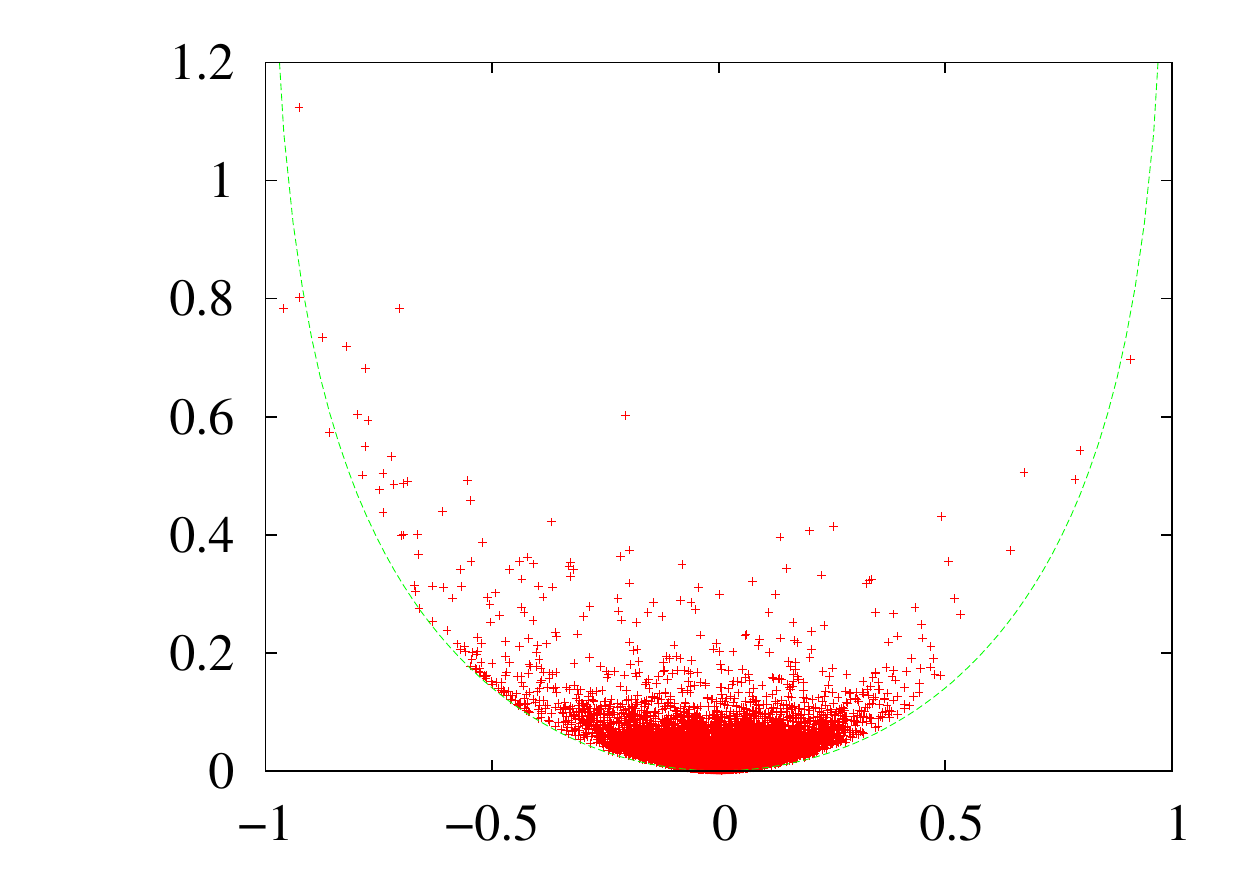}}\\
\caption{Mutual information $MI$ vs. linear correlation coefficient $r$ plots for four selected proteins. The green dashed line is a universal fit for all ten studied proteins and is given by equation \ref{eq:fit}. } 
\label{fig:mi-r}
\end{figure}

As being evident from the full correlation expansion\cite{Matsuda2000}, mutual information is an inherent component of entropy, thus is intimately related to free energy at a given temperature. Therefore, utility of mutual information to characterize dynamical correlations makes more energetic sense than both linear correlations or the generalized correlation\cite{Lange2006}. While this fact has been well recognized
, it remains unclear how linear correlations relate to mutual information, and consequently free energy in proteins. To elucidate this issue, we calculated second order mutual information ($MI$) and linear correlation coefficients $r$ 
 for all pairs of backbone dihedrals $\phi$ and $\psi$ for ten protein molecules. $MI$ vs. $r$ plots of four proteins were presented in Fig. \ref{fig:mi-r}
. Contour lines of these scatter plots approximately reflect relationship 
between $r$ and maximum possible $MI$ (denoted $MPMI_r$ here after) engendered by corresponding linear correlations (as there is always possibility of non-linear correlations between any given BTPs). It is found that contour lines are essentially the same regardless of the identity of proteins, and may be reasonably well-fit (Fig. \ref{fig:mi-r} 
) with the following function.
\begin{equation}
MI = \frac{-ln(1-r)ln(1+r)}{2}
\label{eq:fit}
\end{equation}
Meanwhile, data points locate above and far from contour lines indicate that significant non-linear correlations exist for corresponding BTPs. Again, note that $MI$ is linearly related to entropy by the Boltzmann constant. Therefore, for points fall on equation \ref{eq:fit}, entropic cost for initial increase of $r$ from $0.1$ to $0.4$ is around $0.08k_B$, while a further increase of $r$ from $0.4$ to $0.7$ corresponds to approximately $0.23k_B$. Additionally, while the theoretical range of $MI$ goes from $0$ to $\infty$, thermodynamics dictates that we will not observe huge values in practical biomolecular systems, which operate under ambient conditions. Indeed, as shown in Fig. \ref{fig:mi-r} 
, the maximum $MI$ we observed is less than $1.3$ and $MI$ value beyond $1.0$ is extremely rare. Therefore, utility of $MI$ to characterize dynamical correlations provides both practical convenience and physical intuition. 

\subsection{Sequential distribution of BTP correlations}
\begin{figure}
\centering 
\subfloat[1rgh]{\includegraphics[width=3.5in]{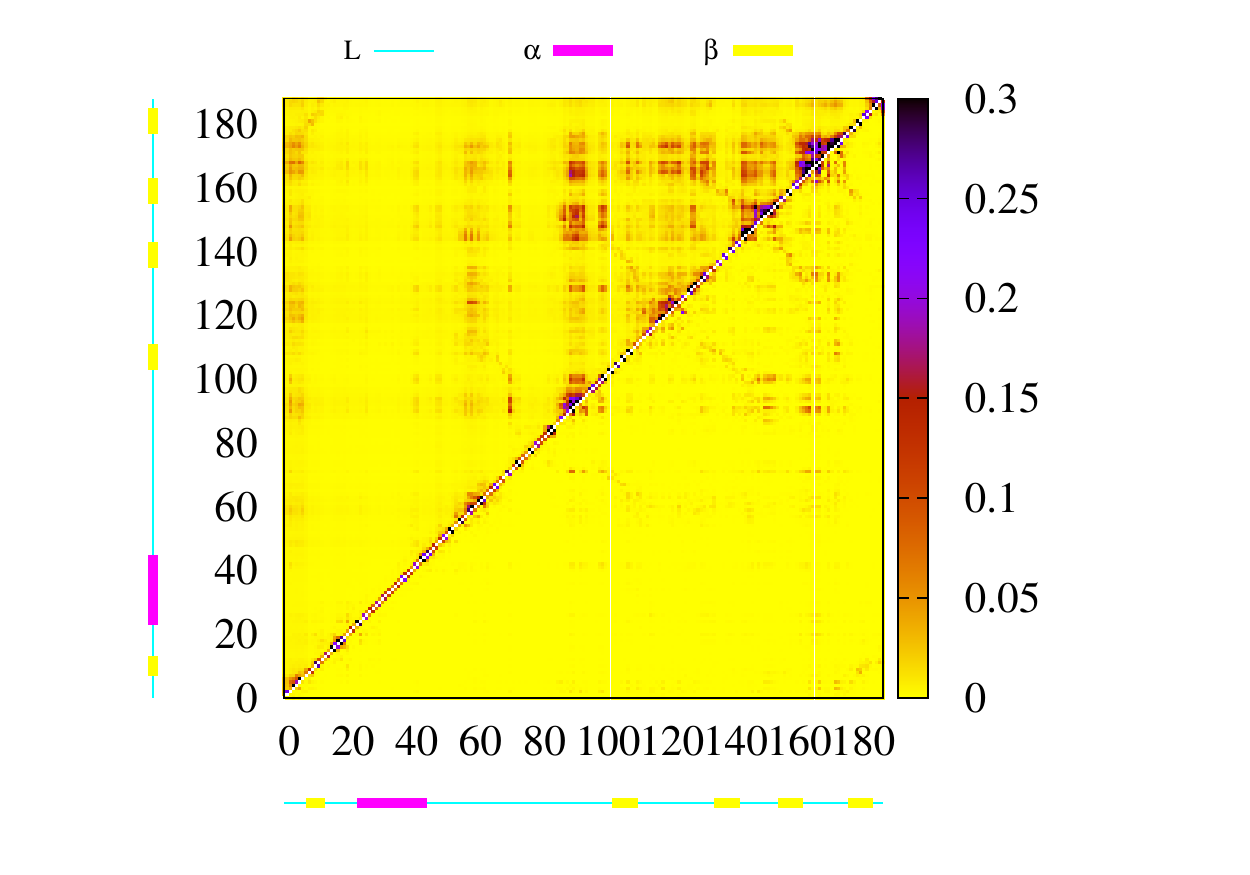}}
\subfloat[7rsa]{\includegraphics[width=3.5in]{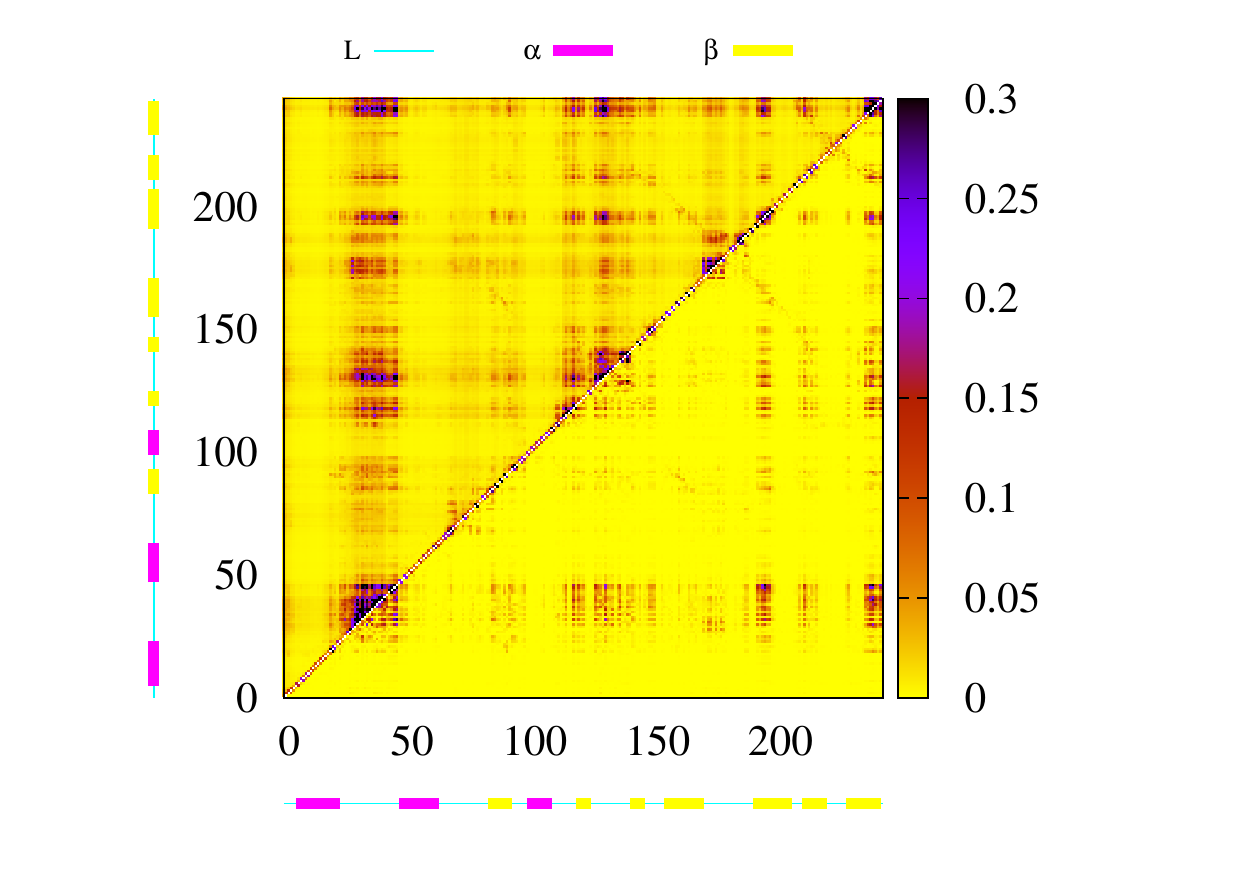}}\\
\subfloat[cdk2]{\includegraphics[width=3.5in]{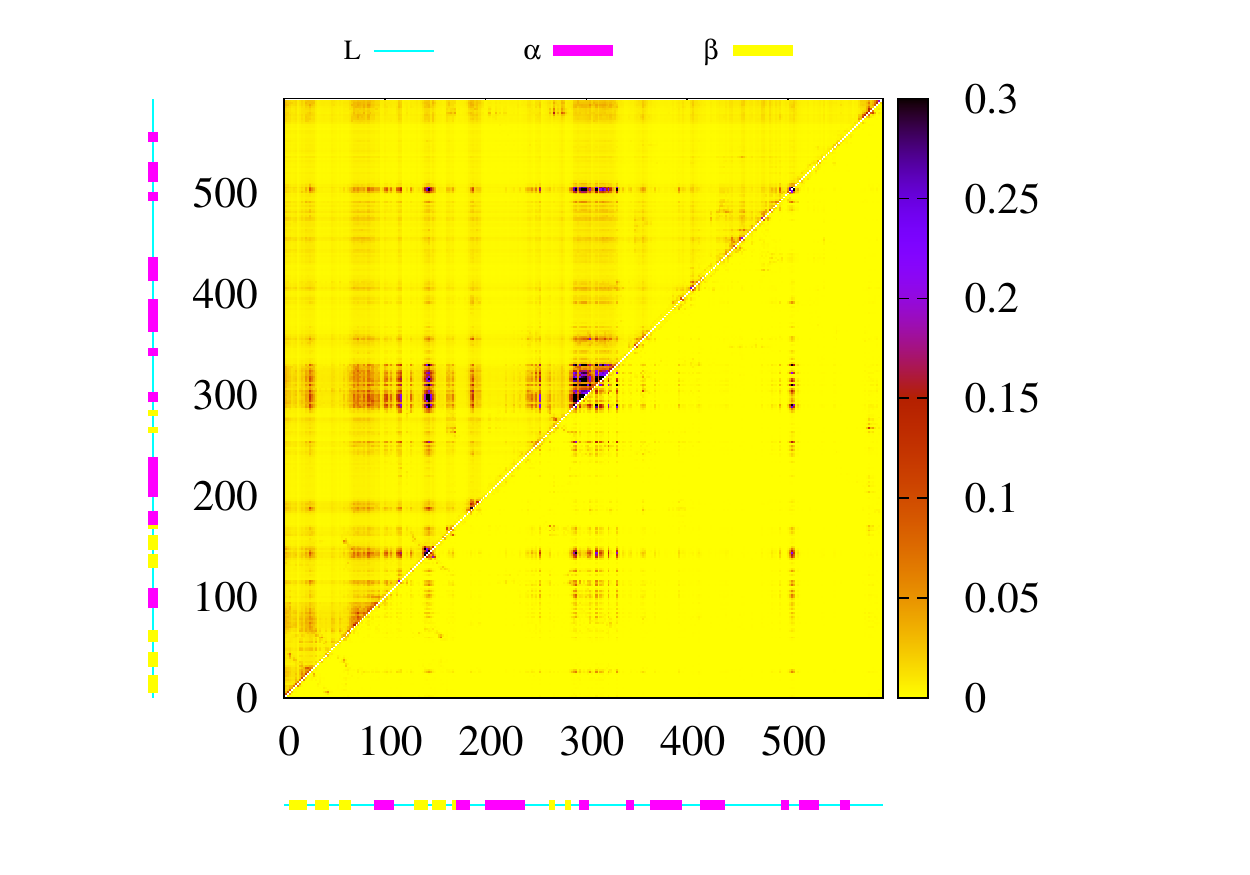}}
\subfloat[lyzm]{\includegraphics[width=3.5in]{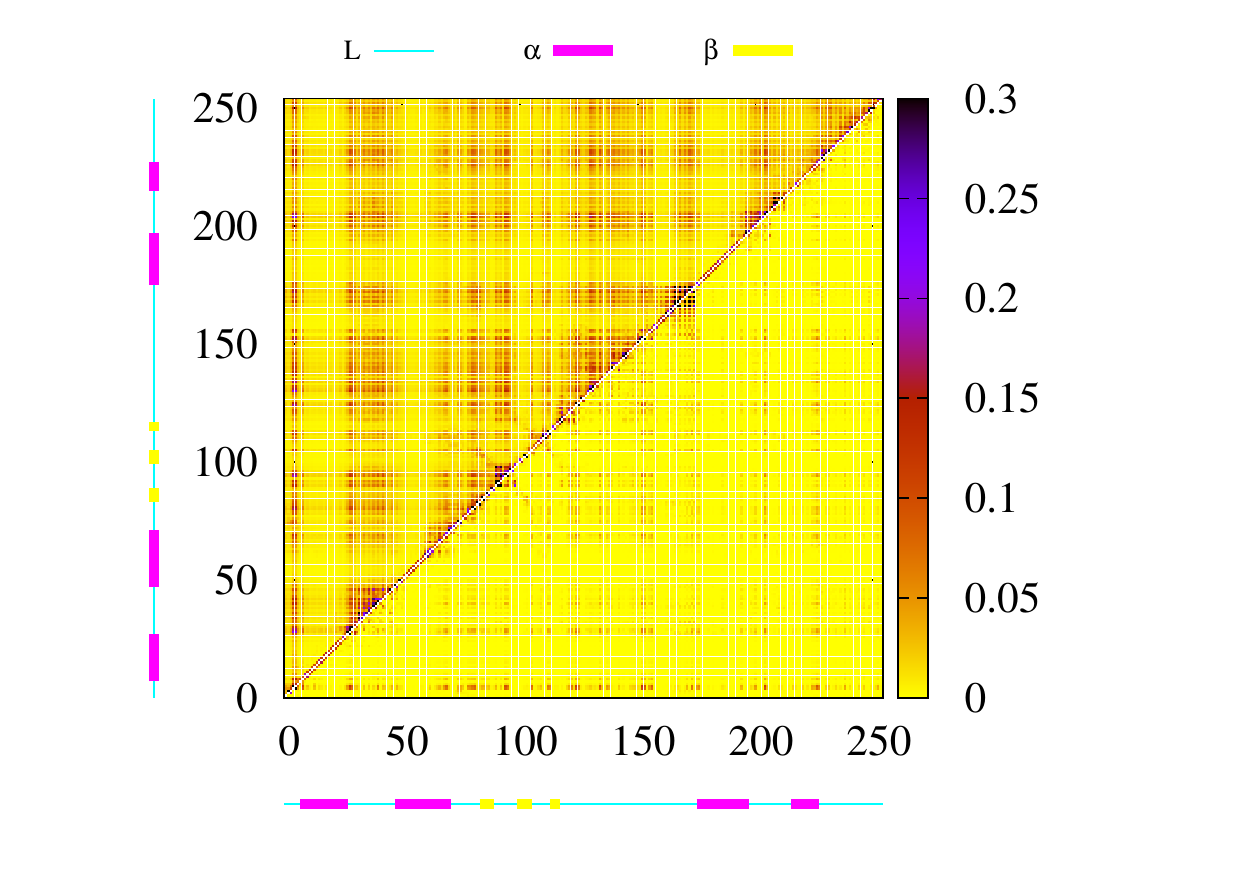}}\\
\caption{Correlation matrices of four selected proteins. For each protein, the full mutual information (MI) is shown in upper-left triangle, and the $MPMI_r$ transformed from linear correlation coefficient $r$ is shown in lower-right triangle. The numbers in both horizontal and vertical axis are indices of backbone torsions, which run from N-terminus to C-Terminus. Strength of correlation is indicated by the color bar to the right side. See Fig. ??? for $MI$ vs. $r$ plots of the remaining six analyzed proteins.} 
\label{fig:cmat}
\end{figure}

To analyze the distribution of both linear correlations and $MI$ of BTPs in primary sequence space, a correlation matrix was created for each analyzed protein and presented in Fig. \ref{fig:cmat} 
For convenience of presentation on the same matrix, $r$ was first transformed into $MPMI_r$ by utilizing equation (\ref{eq:fit}). For sequentially long-ranged pairs (off diagonal points in correlation matrices), the full $MI$ (presented in left-upper half matrices), which includes both linear and nonlinear contributions, is significantly larger than $MPMI_r$ in most of proteins analyzed. By limiting the range of $MI$ (and $MPMI_r$) to $[0, 0.3]$, correlations of BTPs formed by immediate neighboring torsions in sequence were effectively excluded for a better view of correlation patterns elsewhere. 
The observation suggests that, when compared with linear correlations, non-linear contributions are increasingly more important over longer distances in the primary sequence. However, the extent of difference between full $MI$ and $MPMI_r$ varies for different proteins (see Fig. \ref{fig:cmat}). Additionally, a common feature shared by all proteins is that significant $MI$ in off-diagonal region is primarily associated with various loop residues (all residues that were in neither an $\alpha$ helix nor a $\beta$ strand were defined as loop residues in this study).

\subsection{Relevance of spatial distances and secondary structures for non-linear BTP correlations}
\begin{figure}
\centering 
\subfloat[1rgh]{\includegraphics[width=2.5in]{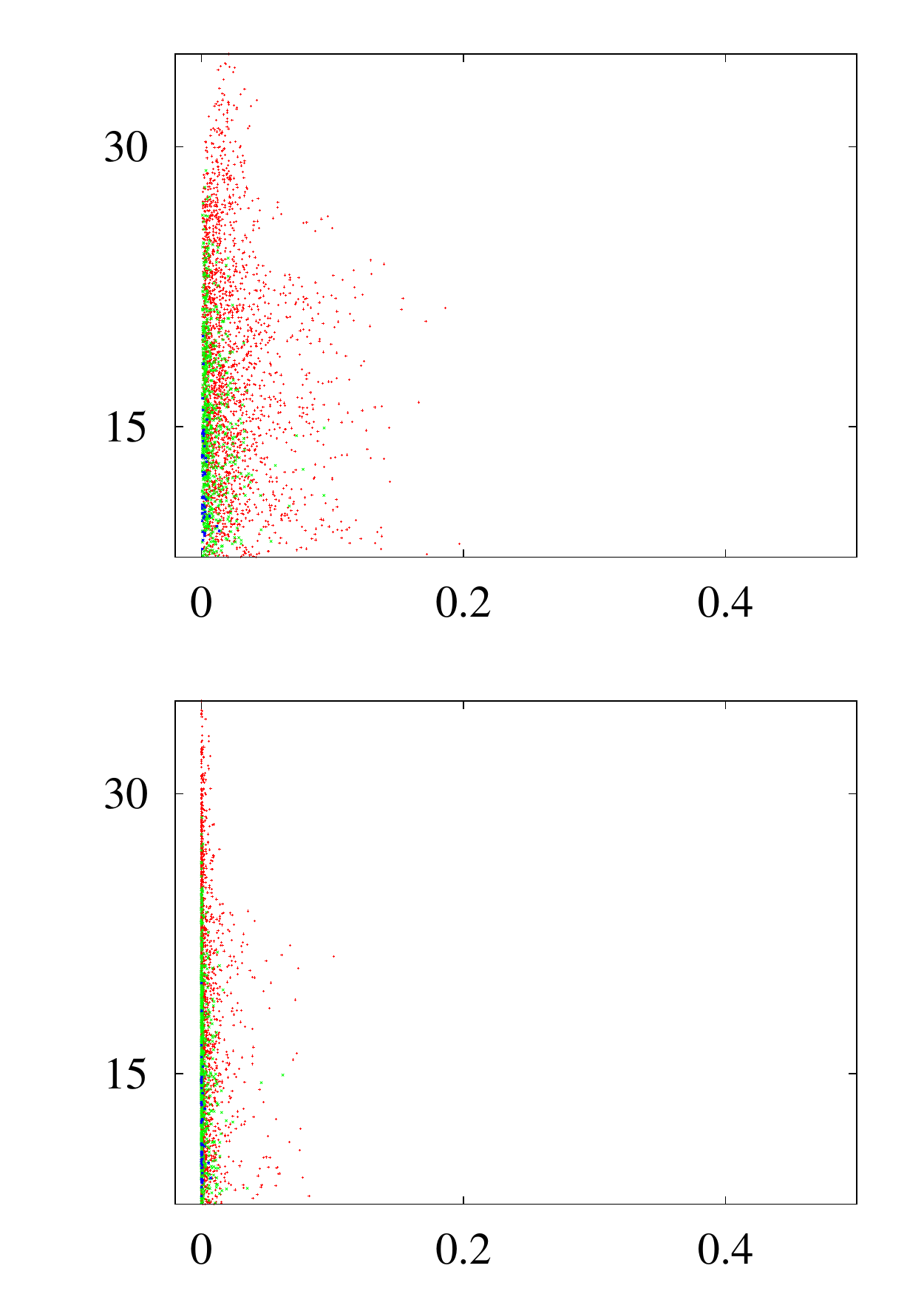}}
\subfloat[7rsa]{\includegraphics[width=2.5in]{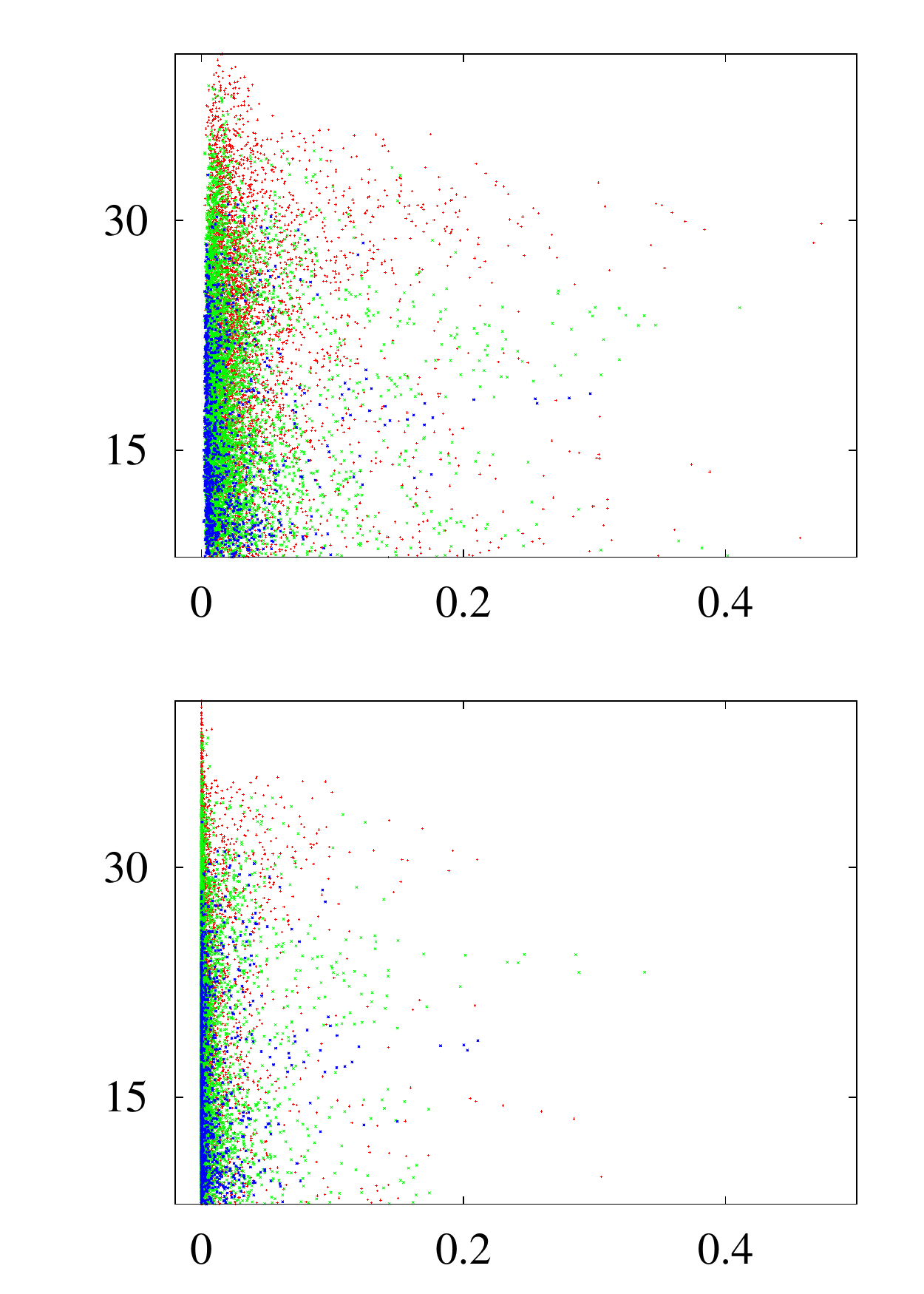}}\\
\subfloat[cdk2]{\includegraphics[width=2.5in]{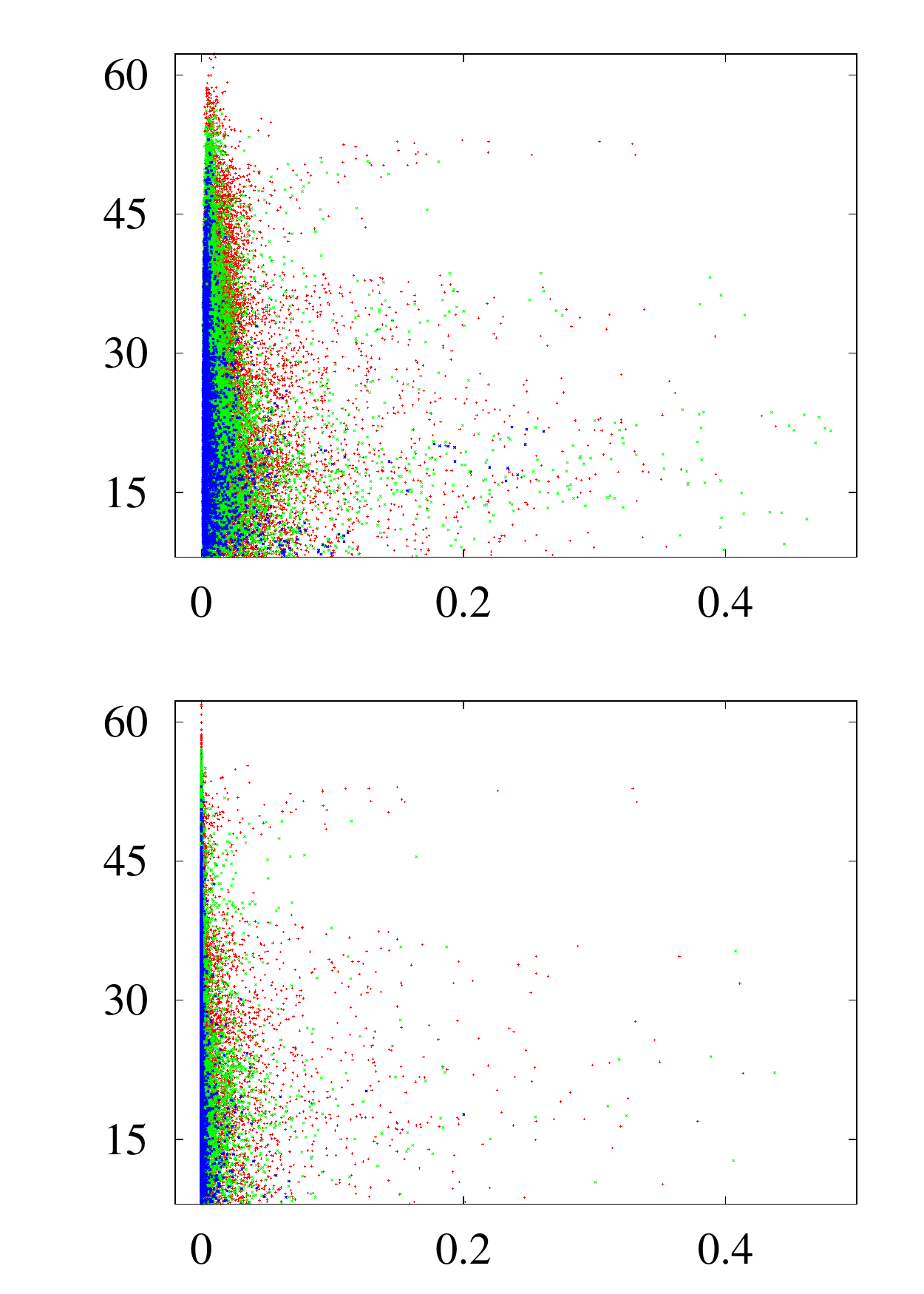}}
\subfloat[lyzm]{\includegraphics[width=2.5in]{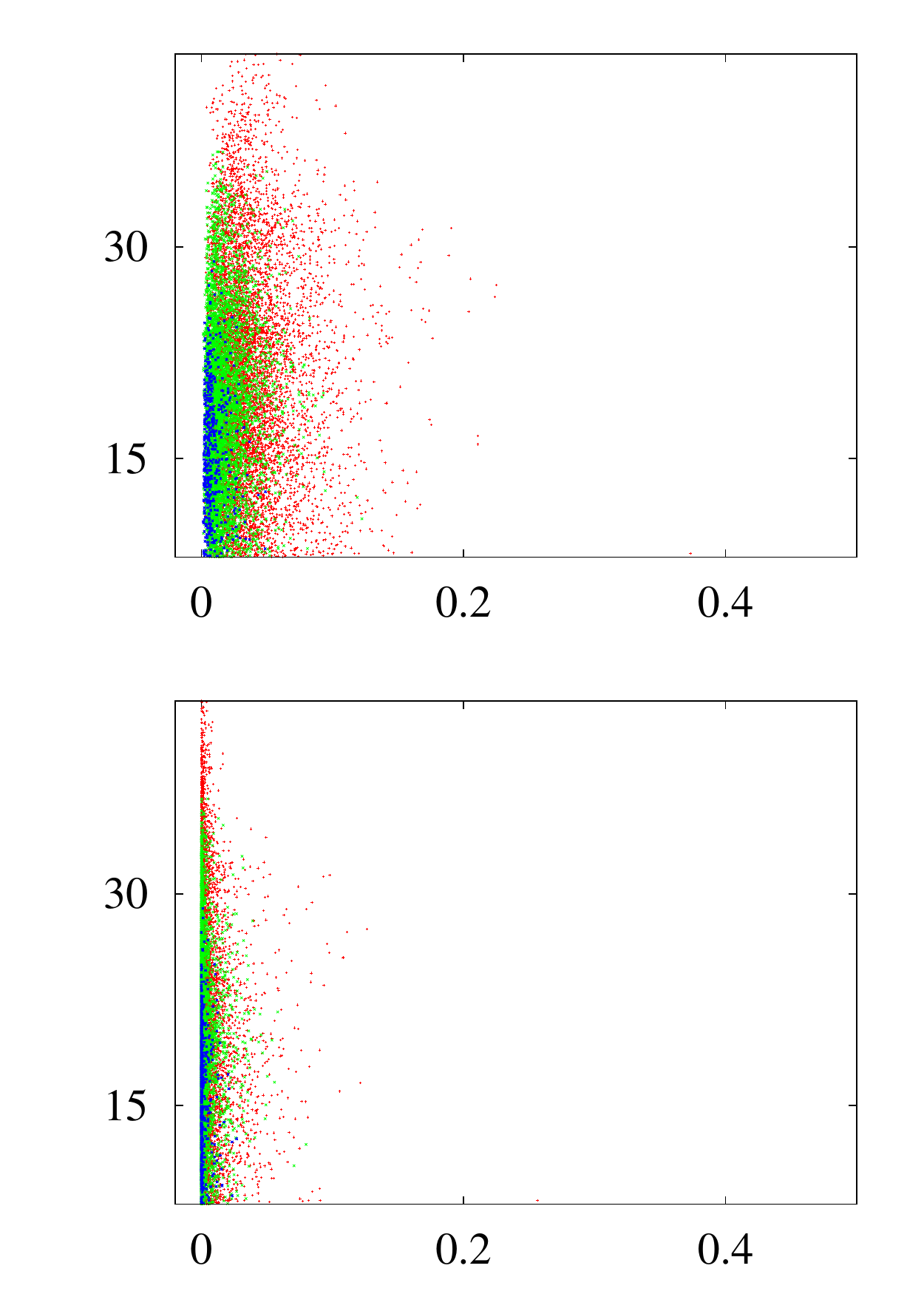}}\\
\caption{$MI$ (top panels) and $MPMI_r$ (bottom panels) for for selected proteins. $\alpha/\beta$-$\alpha/\beta$ BTPs are shown in blue, $\alpha$-$L$ BTPs are shown in green, and $L$-$L$ BTPs are shown in red. } 
\label{fig:dist}
\end{figure}

In three dimensional protein structures, large distances in primary sequence may correspond to either long or short distances in space. Correlations caused by physical adjacency are trivially expected in condensed phases. In practice, what we care most are spatially long-ranged (SLR) correlations due to their potential participation in functionally important allosteric interactions. To analyze spatial variance of BTP correlations, the calculated $MI$ and $MPMI_r$ were plotted with respect to spatial distances as shown in Fig. \ref{fig:dist} 
Two major consistent features were observed in all of studied proteins. Firstly, SLR correlations have significant nonlinear contributions since $MI$ are generally larger than corresponding $MPMI_r$, especially at large spatial distances. Secondly, loop-loop ($L$-$L$) BTPs exhibit the most and the largest, $\alpha$ helix and $\beta$ strand ($\alpha$/$\beta$-$\alpha$/$\beta$) BTPs have the least and the smallest, and $\alpha$/$\beta$-loop ($\alpha$/$\beta$-L) BTPs manifest intermediate SLR correlations. Regarding the second feature, significant variation was observed among different proteins (Fig. \ref{fig:dist}). 

\begin{figure}
\centering 
\subfloat[1rgh]{\includegraphics[width=3.in]{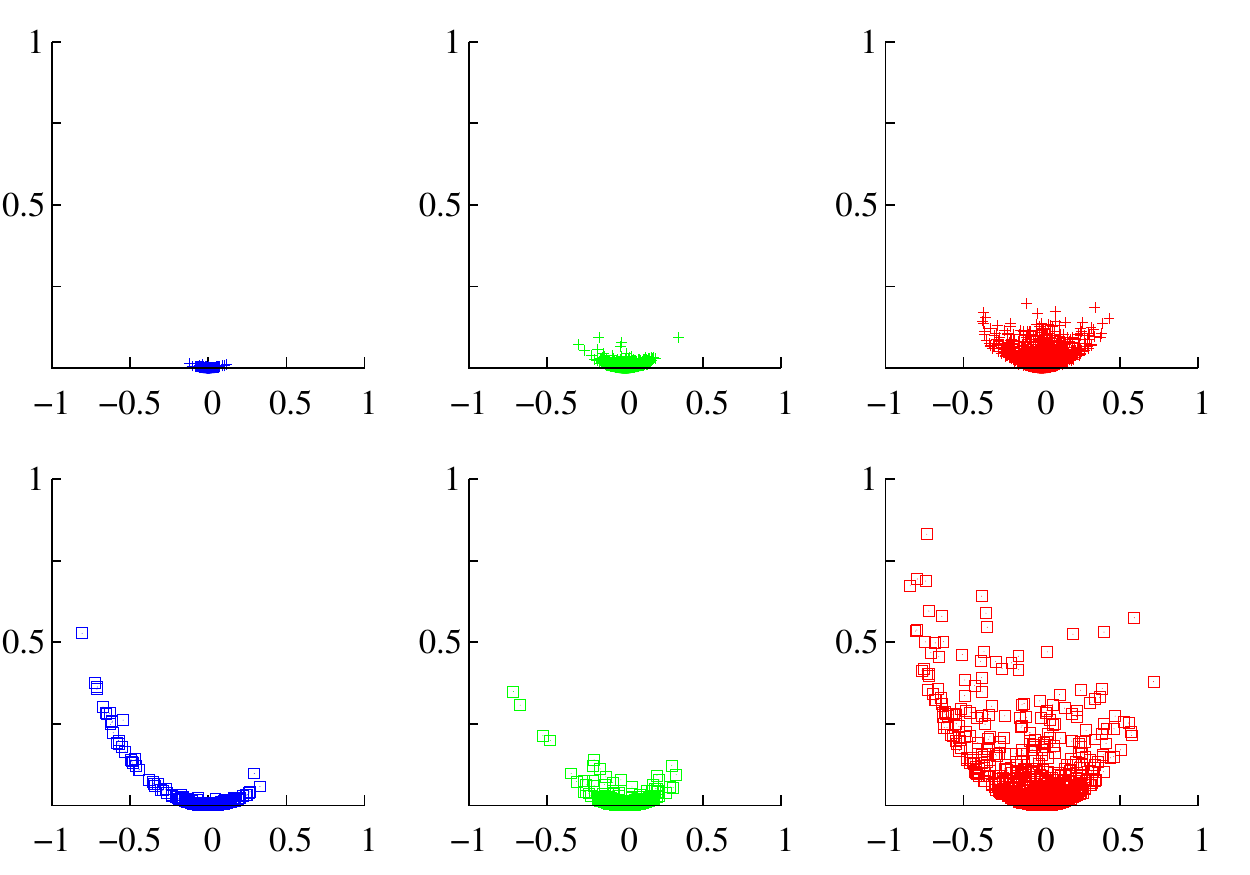}}
\subfloat[7rsa]{\includegraphics[width=3.in]{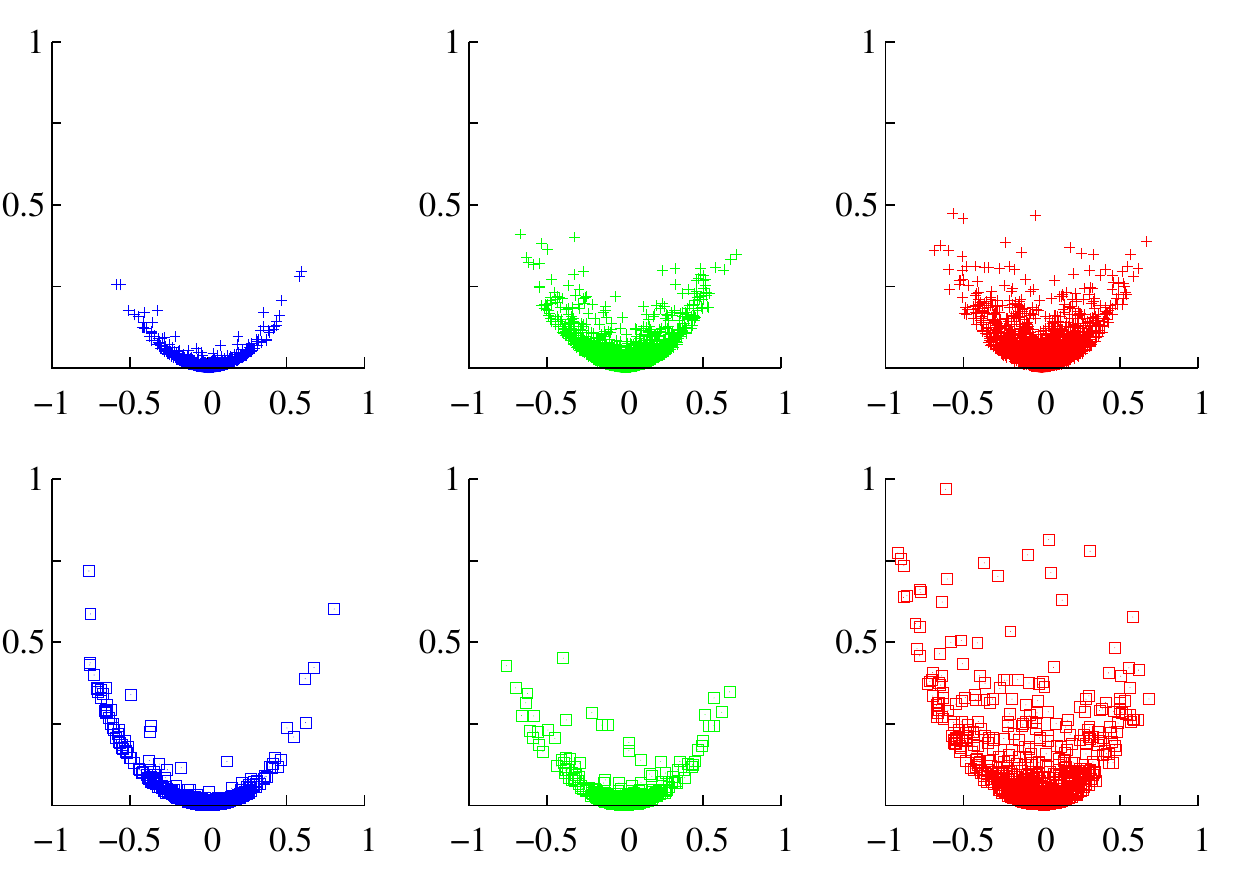}}\\
\subfloat[cdk2]{\includegraphics[width=3.in]{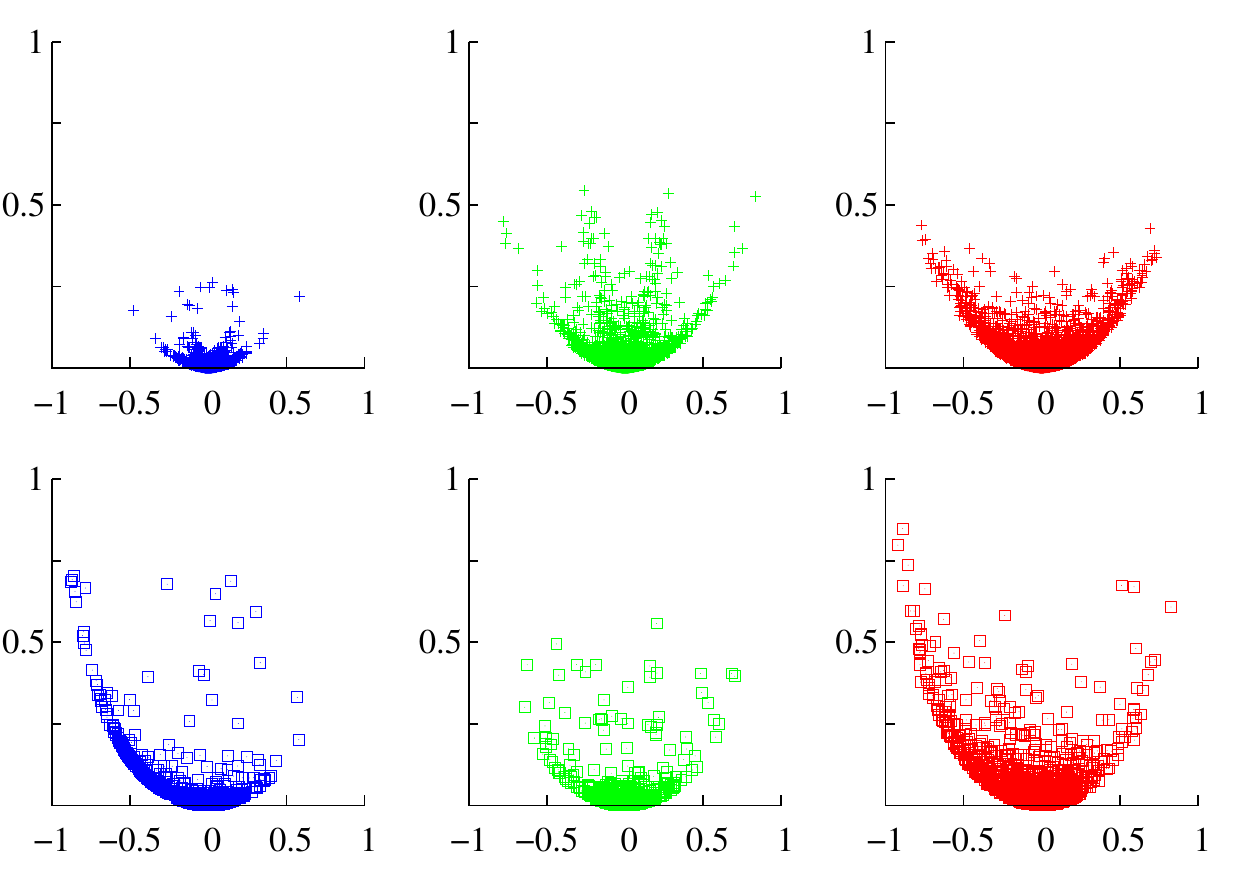}}
\subfloat[lyzm]{\includegraphics[width=3.in]{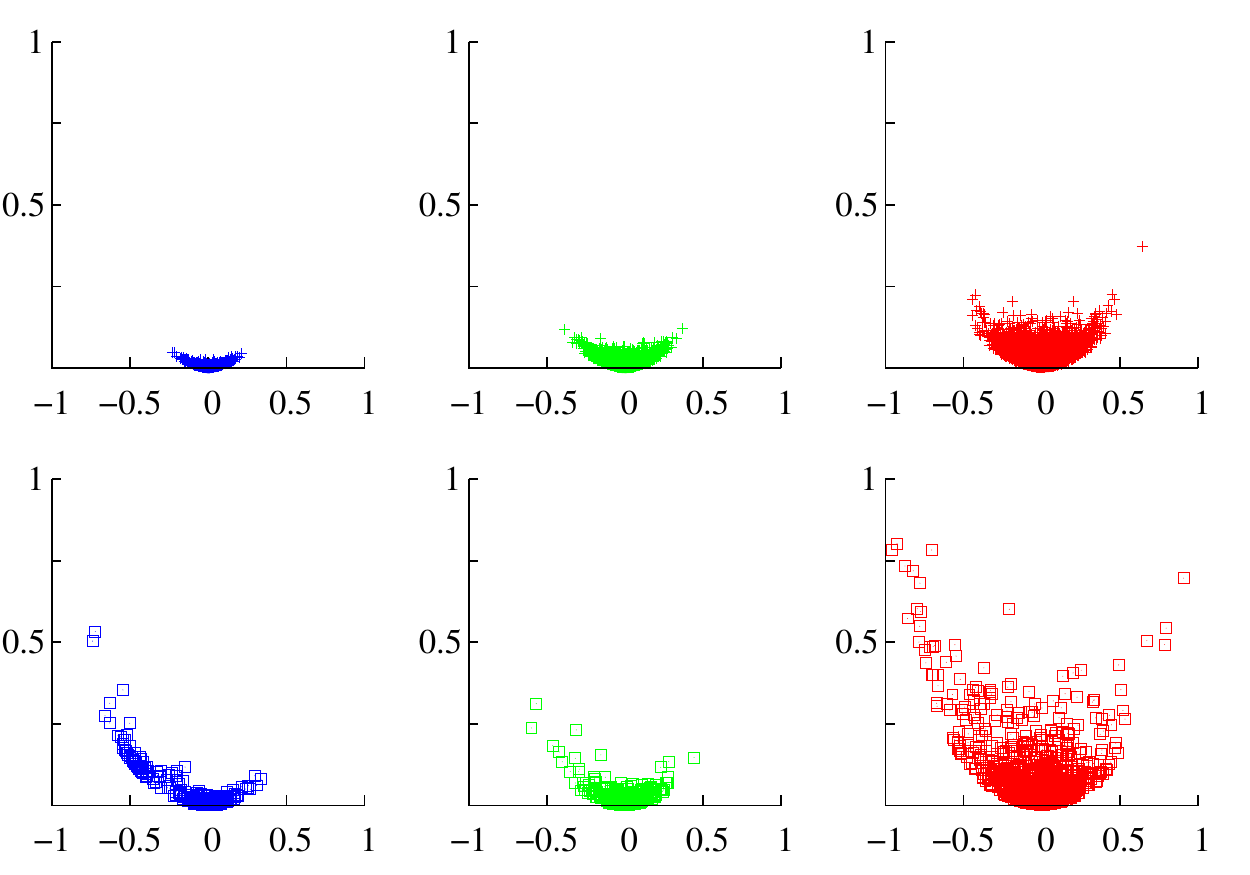}}\\
\caption{$MI$ vs. $r$ plots for Local (distances between two participating torsions of a BTP is smaller than or equal to 8 $\AA$, crosses in top panels) and long-range (otherwise, squares in bottom panels) different types of BTPs. $\alpha/\beta$-$\alpha/\beta$ BTPs are shown in blue, $\alpha$-$L$ BTPs are shown in green, and $L$-$L$ BTPs are shown in red. } 
\label{fig:distss}
\end{figure}

Qualitatively, correlation matrices for studied proteins (see Fig. \ref{fig:cmat}) suggest that for those relatively strong sequentially non-local correlations (off-diagonal region), nonlinear contributions are significant. Similarly, distance vs. correlation plots in Fig. \ref{fig:dist} indicate that SLR correlations have significant nonlinear contributions and this is especially true for some $L$-$L$ BTPs. To further clarify relative importance of nonlinear correlations for different types of BTPs (i.e. $L$-$L$, $\alpha$/$\beta$-$L$ and $\alpha$/$\beta$-$\alpha$/$\beta$) and different spatial distances, we constructed $MI$ vs. $r$ plots for spatially local (with distances equal or smaller than 8 angstroms) and non-local (otherwise) for each type of BTPs and presented the results in Fig. \ref{fig:distss}. For most proteins, $\alpha$/$\beta$-$\alpha$/$\beta$ BTPs exhibit mainly linear correlations with the overwhelming majority of data points fall on the indicated contour line, $L$-$L$ BTPs have the most number of data points exhibit significant nonlinear correlations, and $\alpha$/$\beta$-$L$ BTPs stays in between. Spatial locality, while makes decisive differences in correlation strength, plays a unimportant role in relative significance of linear and non-linear contributions among different types of BTPs. 
It is important to note that, in all studied proteins, the majority BTPs fall on or locate closely to the contour line in $MI$ vs. $r$ plots (Fig. \ref{fig:distss}) regardless of specific BTP types. Therefore, linear correlation contributes dominantly for most of BTPs iirespect of the specific secondary structures in which the participating torsion locate. It is only that $L$-$L$ BTPs are the mostly likely, and $\alpha$/$\beta$-$\alpha$/$\beta$ BTPs are the least likely to have significant nonlinear contributions to their correlations, with $\alpha$/$\beta$-$L$ BTPs being the intermediate scenario in this regard. 

\subsection{Torsional state transitions and non-linear BTP correlations}
Based on observations mentioned above, we were quite confident that neither spatial distances nor the specific identity of belonging secondary structures \emph{per se} is a necessary factor for significant nonlinear contribution in BTP correlations. Instead, it should be some other property that is most likely to associate with loop residues and is least likely to associate with residues in stable secondary structures. For backbone torsions in loops, one outstanding feature is significantly higher (relative to those located in $\alpha$ helices and $\beta$ strands) probability of having multiple torsional states on various time scales. In contrast, most backbone dihedrals in stable secondary structures stay in one specific torsional state for native proteins. To test for necessity of torsional state transitions in nonlinear contributions to BTP correlations, we calculated distributions of all $\phi$s and $\psi$s for each of studied proteins and searched for torsional state transitions according to the specified rule 
Indeed, for points above the contour line (as specified by equation \ref{eq:fit}), torsional states transitions were observed for all BTPs. For points fall on or locate very close to the line, only a small fraction of BTPs have participating torsions experiencing torsional state transitions.

\begin{figure}
\centering 
\subfloat[1rgh-sr]{\includegraphics[width=1.6in]{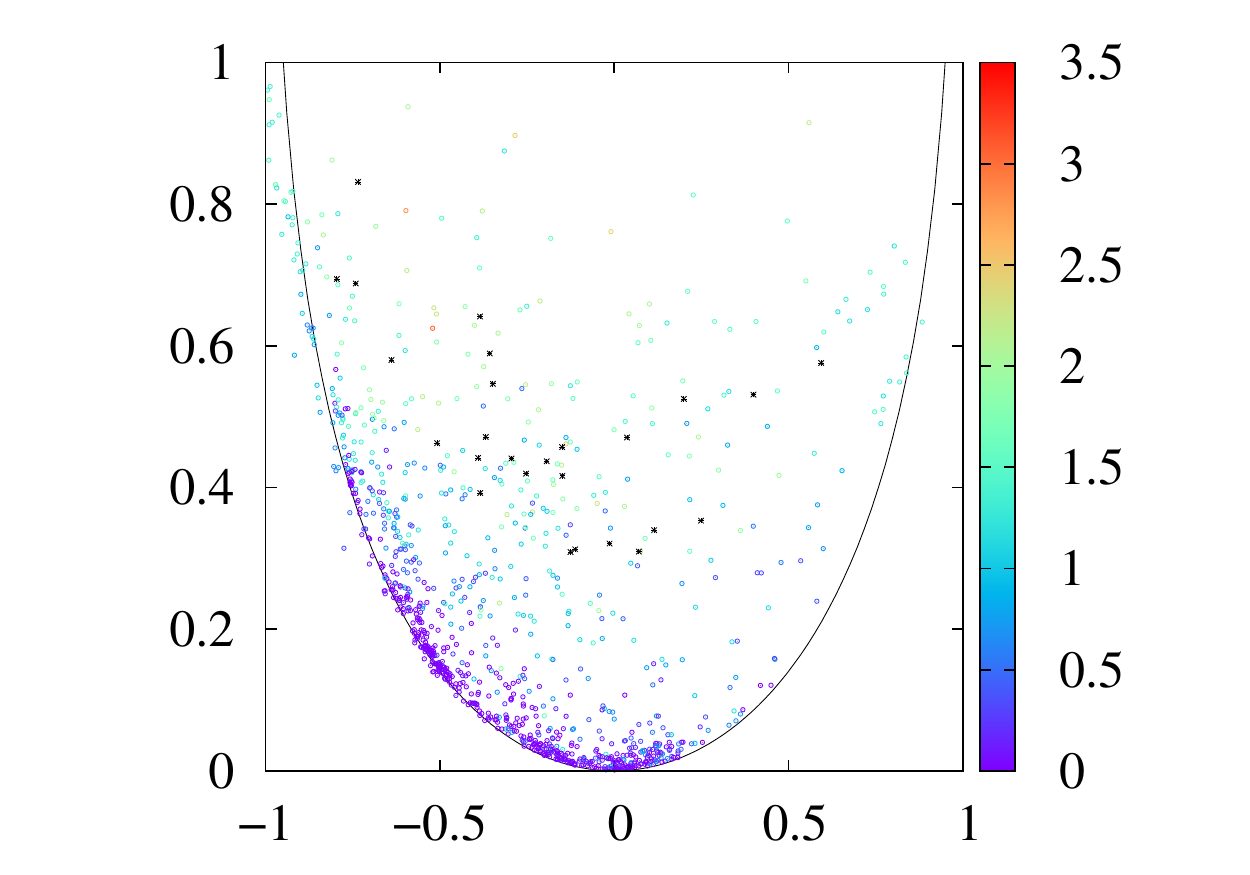}}
\subfloat[7rsa-sr]{\includegraphics[width=1.6in]{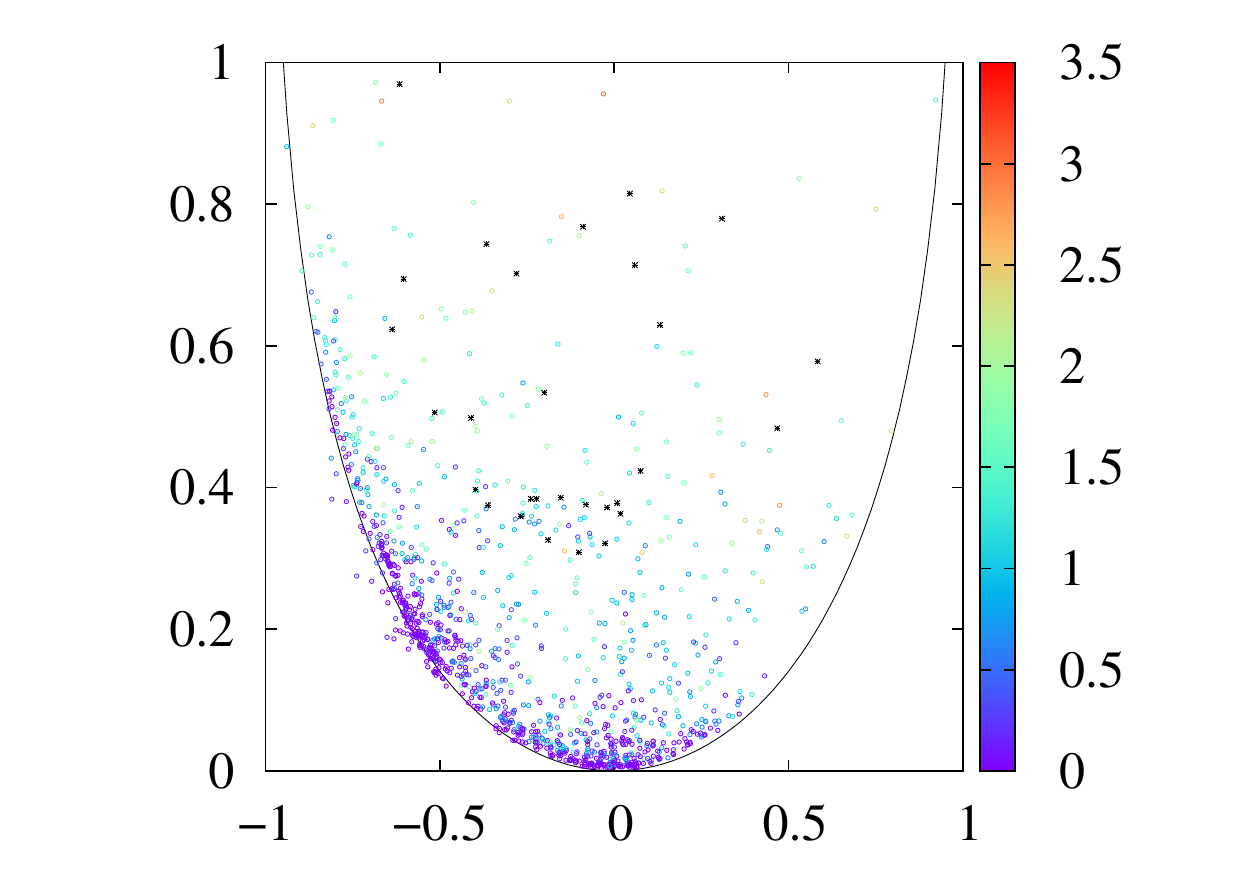}}
\subfloat[cdk2-sr]{\includegraphics[width=1.6in]{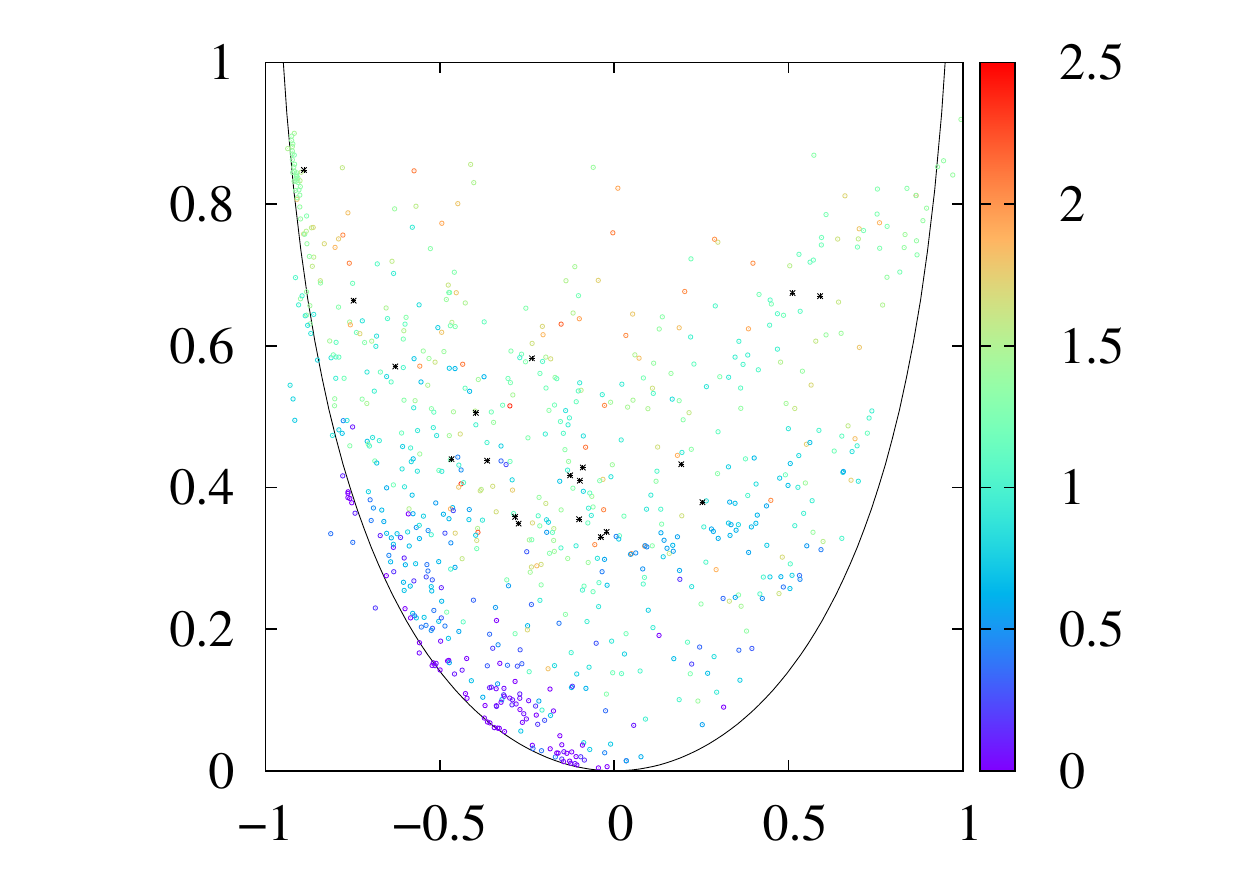}}
\subfloat[lyzm-sr]{\includegraphics[width=1.6in]{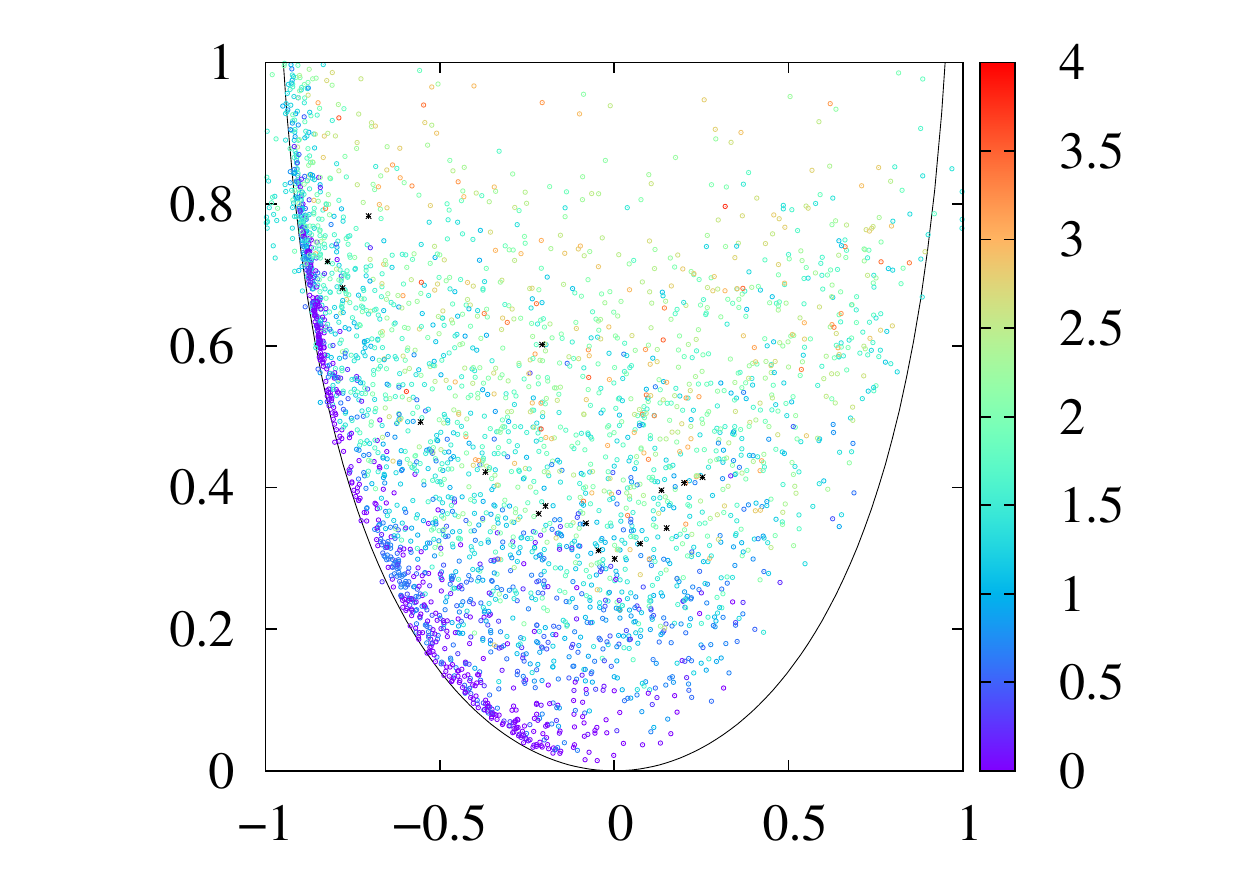}}\\
\subfloat[1rgh-lr]{\includegraphics[width=1.6in]{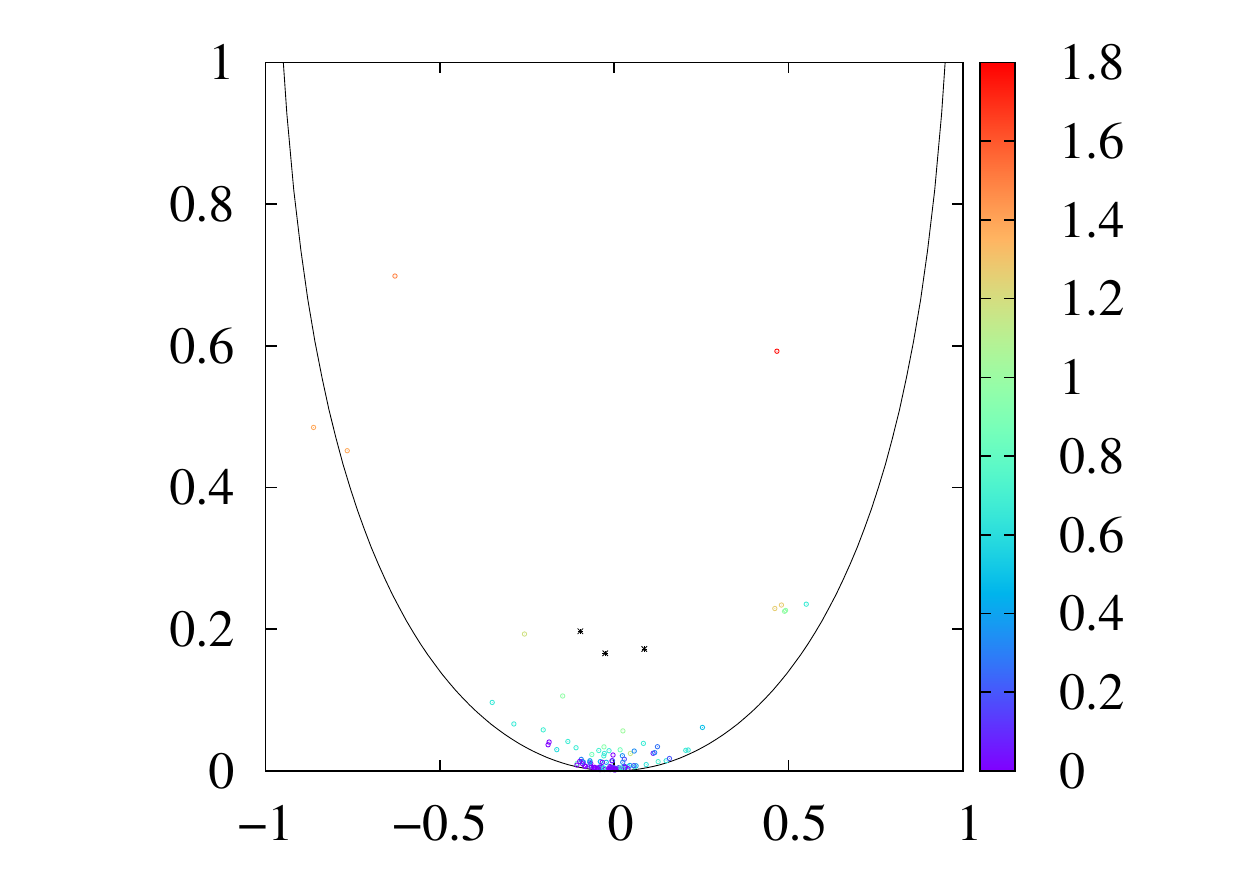}}
\subfloat[7rsa-lr]{\includegraphics[width=1.6in]{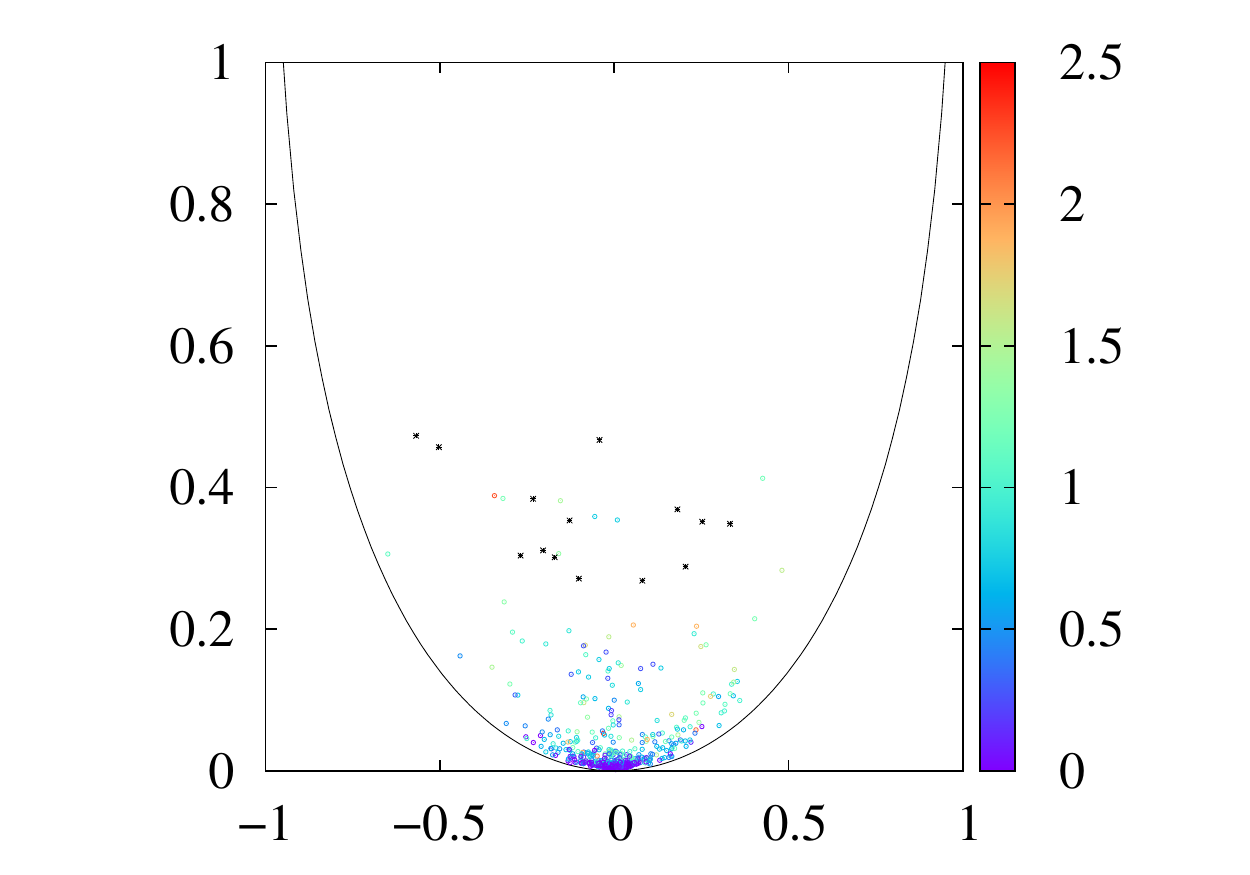}}
\subfloat[cdk2-lr]{\includegraphics[width=1.6in]{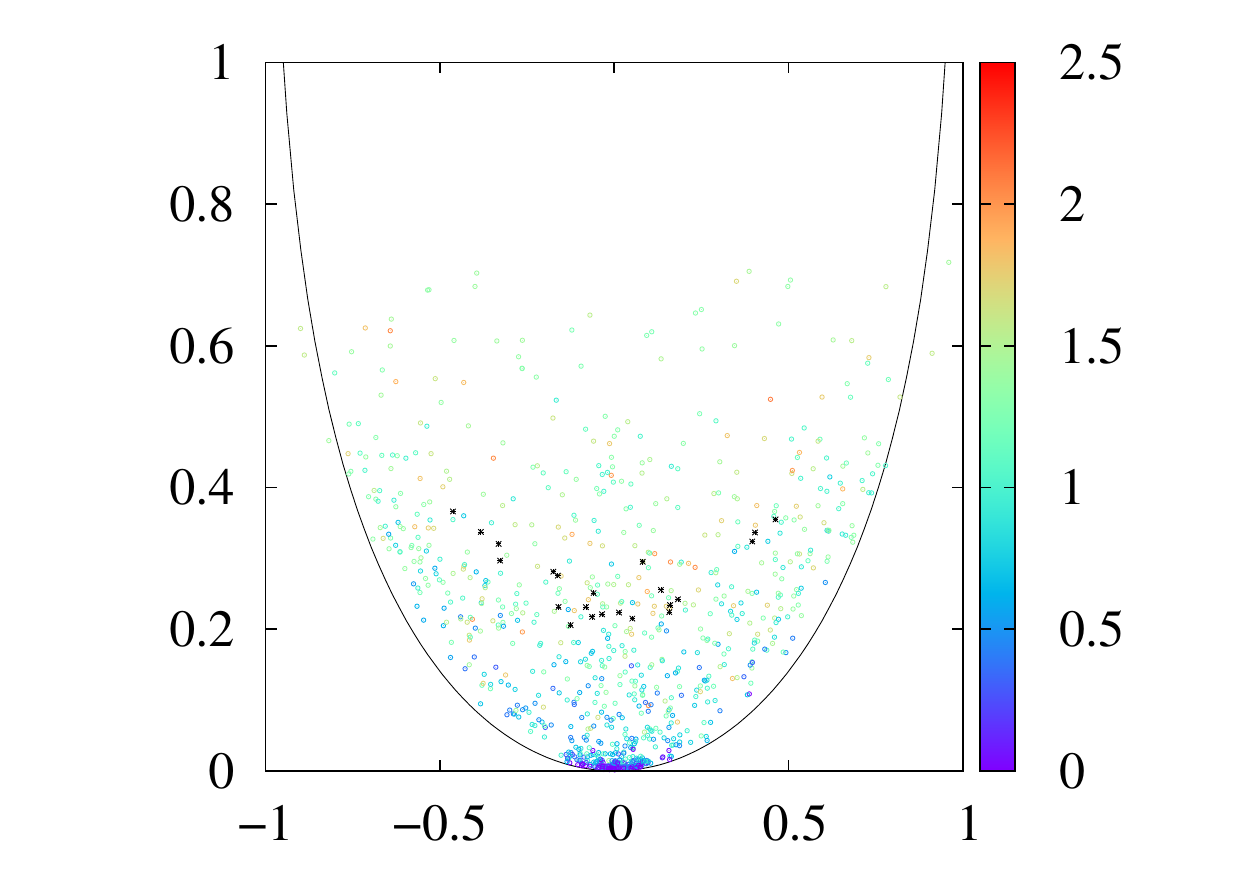}}
\subfloat[lyzm-lr]{\includegraphics[width=1.6in]{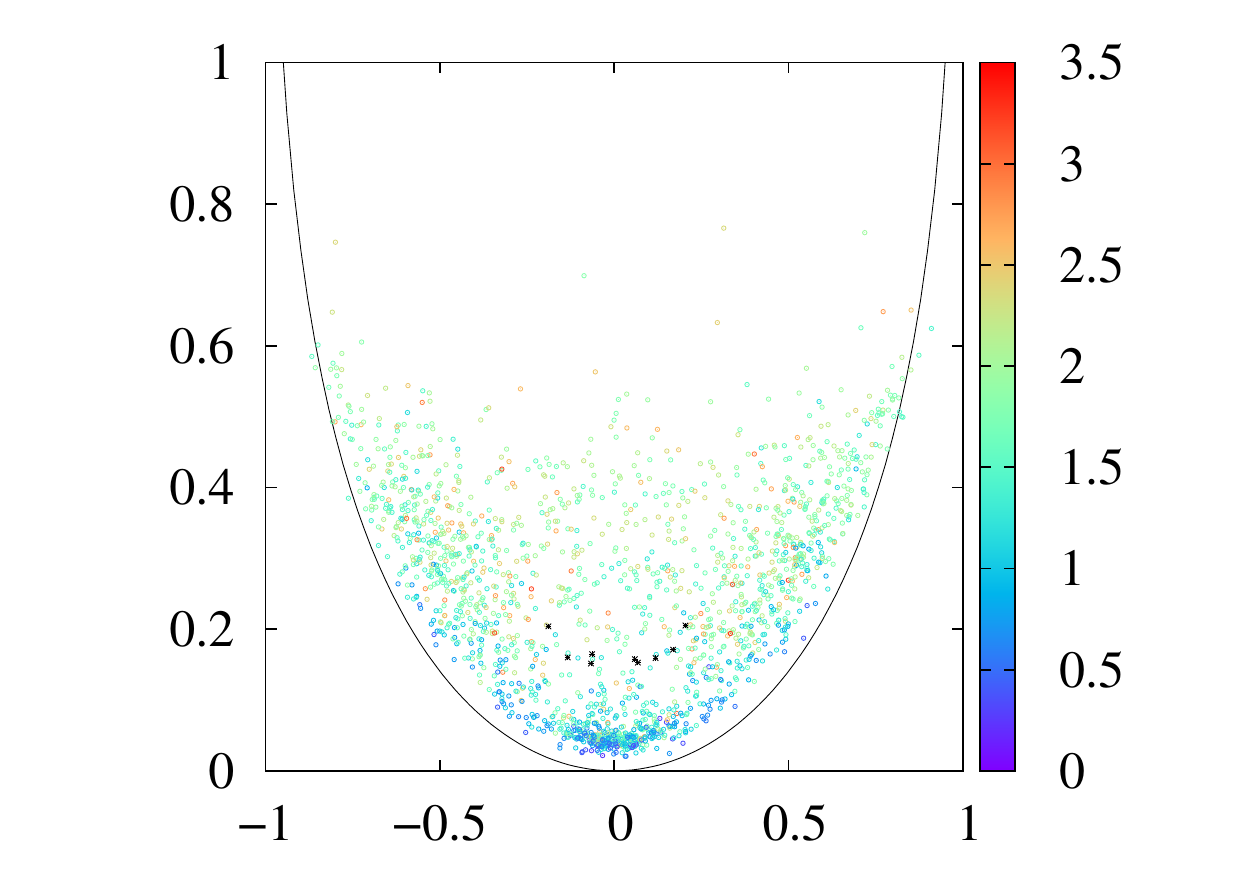}}
\caption{$MI$ vs. $r$ plots for selected BTPs calculated in 40 trajectory subsets for the four selected proteins. Top four panels are for short-ranged BTPs and are indicated by ``-sr'', bottom four panels are for long-ranged BTPs and are indicated by ``-lr''. Each black cross represents a $MI$-$r$ pair of a given BTP in the original collective trajectory set. Each circle represents a $MI$-$r$ pair of a given BTP calculated in one of trajectory subset. Color of circles represents total torsional state entropy of participating torsions for a given BTP in one of trajectory subset. } 
\label{fig:subset}
\end{figure}

While these observations are consistent with the idea that torsional state transitions are necessary for nonlinear BTP correlations, we may not be conclusive. The reason is that for a given protein trajectory set, BTPs that fall on the contour line have different identities and physical environment from those locate above it, and there are other differences between two different BTPs in addition to presence/absence of torsional state transitions for the participating torsions. To resolve these uncertainties, we selected some BTPs that manifest strong nonlinear contributions to pair correlations (locate above and far from contour lines in relevant $MI$-$r$ plots) from each protein and carried out the following analysis. Firstly, we split the original trajectory set into 40 equally sized subsets. Secondly, both $MI$ and $r$ were calculated for each of selected BTP on each of the trajectory subsets. For a given BTP, since torsional state transitions occur on specific time scales, we expect to observe various extent of which in different trajectory subsets. Therefore, by observing the extent of non-linear contributions and torsional DOF distributions from trajectory subsets of the same BTP, we effectively exclude the possibility that observed differences are simply due to the fact of observing different BTPs. $MI$-$r$ plots of selected BTPs obtained from trajectory subsets of the four selected proteins were presented in Fig. \ref{fig:subset}. One common feature of these plots is that relative importance of linear and nonlinear contributions exhibited in trajectory subsets may be widely different from that calculated in the collective set, with majority data points located right onto or in the vicinity of contour lines and remaining ones located far away from contour lines. Further examinations revealed that indeed most data points fall onto contour lines correspond to negligible torsional state transitions as reflected by small torsional state entropy, and data points locate above the contour line exclusively correspond to non-zero torsional state entropy, which demonstrate existence of torsional state transitions of participating torsions. This is consistent with expectation that torsional state transitions generally occur on relatively longer time scales and are rare events on time scale of snapshots recording ($ps$), and therefore was not observed in many trajectory subsets, for which linear correlation dominates. Additionally, for spatially local BTPs (Fig. \ref{fig:subset}abcd), a large fraction of points fall onto or locate close to contour lines exhibit significant linear correlations, while most of data points fall onto or near contour lines have weak linear correlations for SLR BTPs. These observations further suggest that it is quite difficult for linear correlations to propagate over long distances spatially in proteins. 

\subsection{Non-linear correlations and heterogeneous linear correlations}
\begin{figure}
\centering 
\subfloat[]{\includegraphics[width=1.7in]{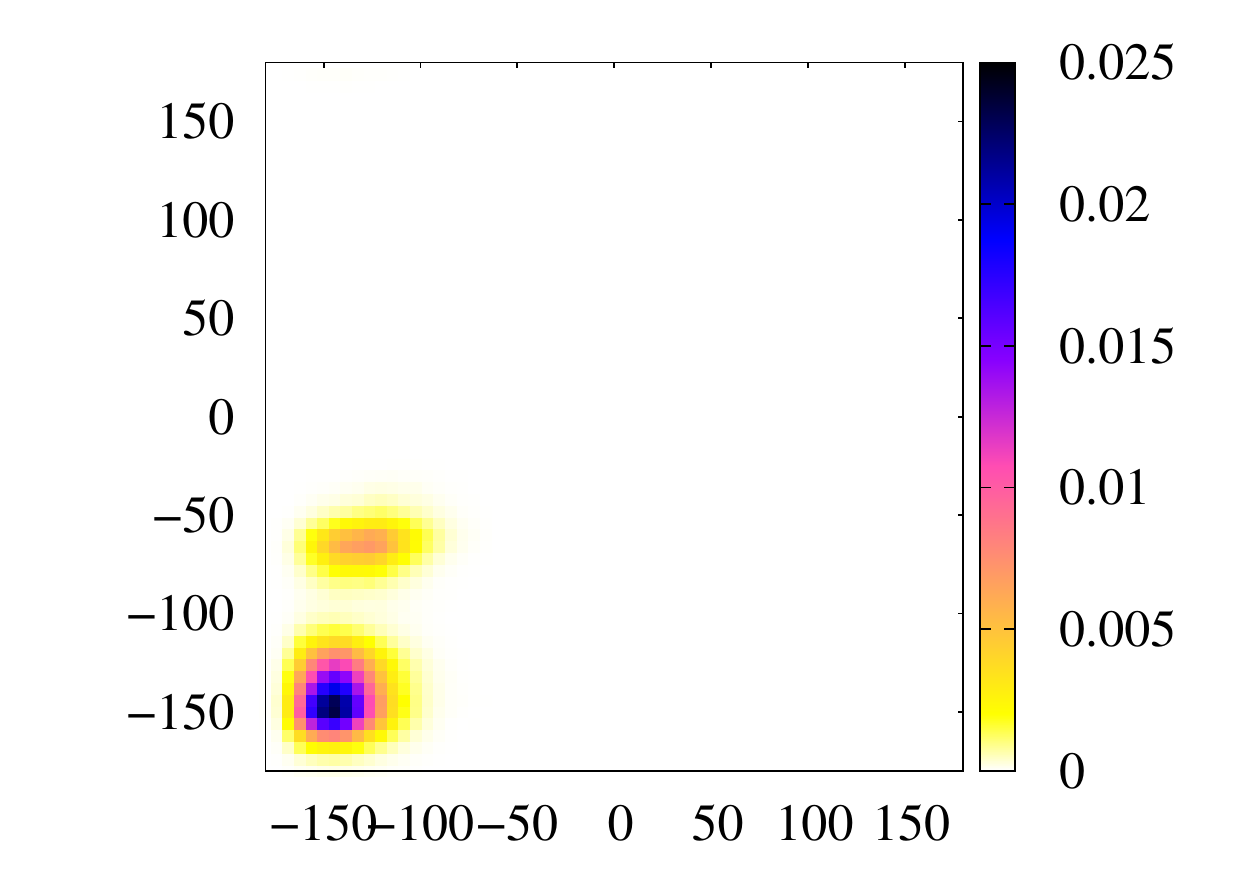}}
\subfloat[]{\includegraphics[width=1.7in]{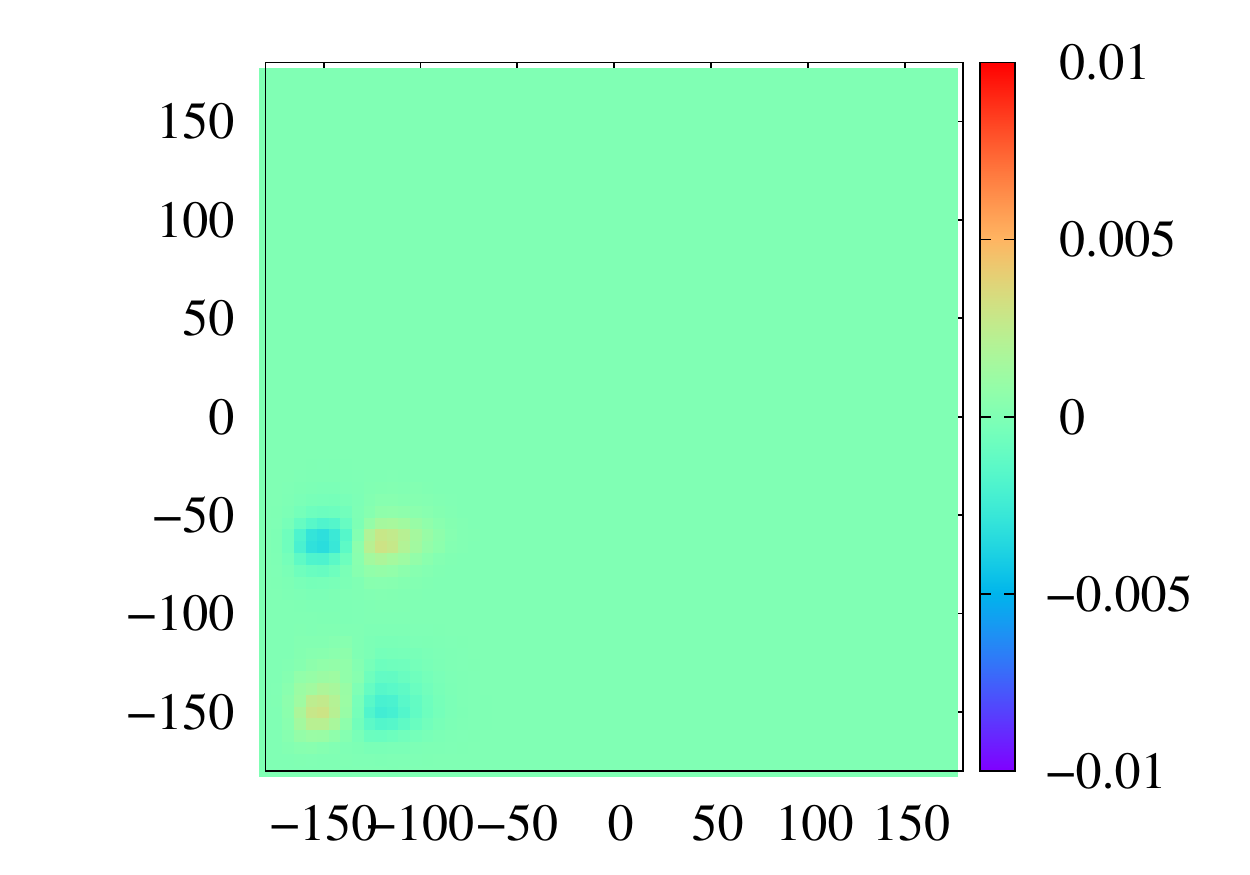}}\\
\subfloat[]{\includegraphics[width=1.7in]{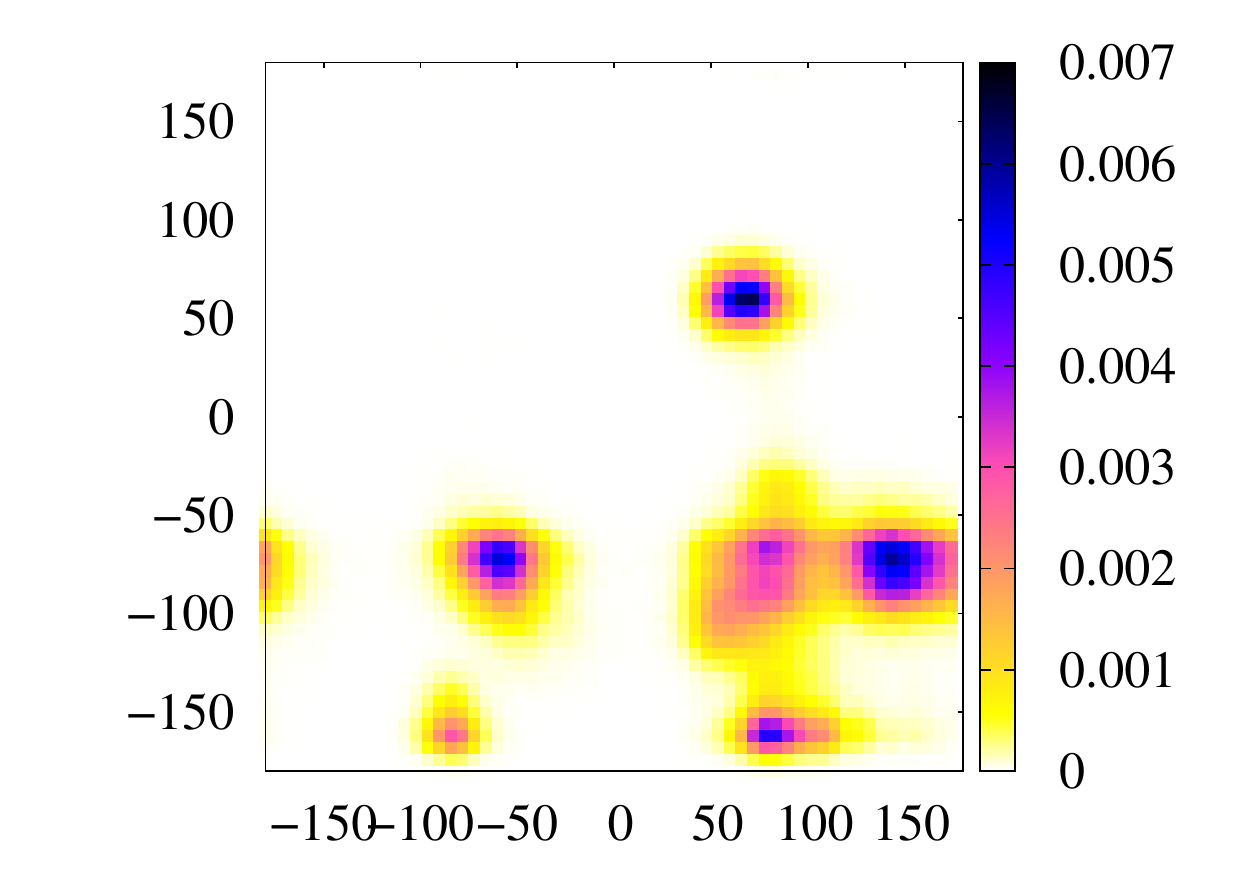}}
\subfloat[]{\includegraphics[width=1.7in]{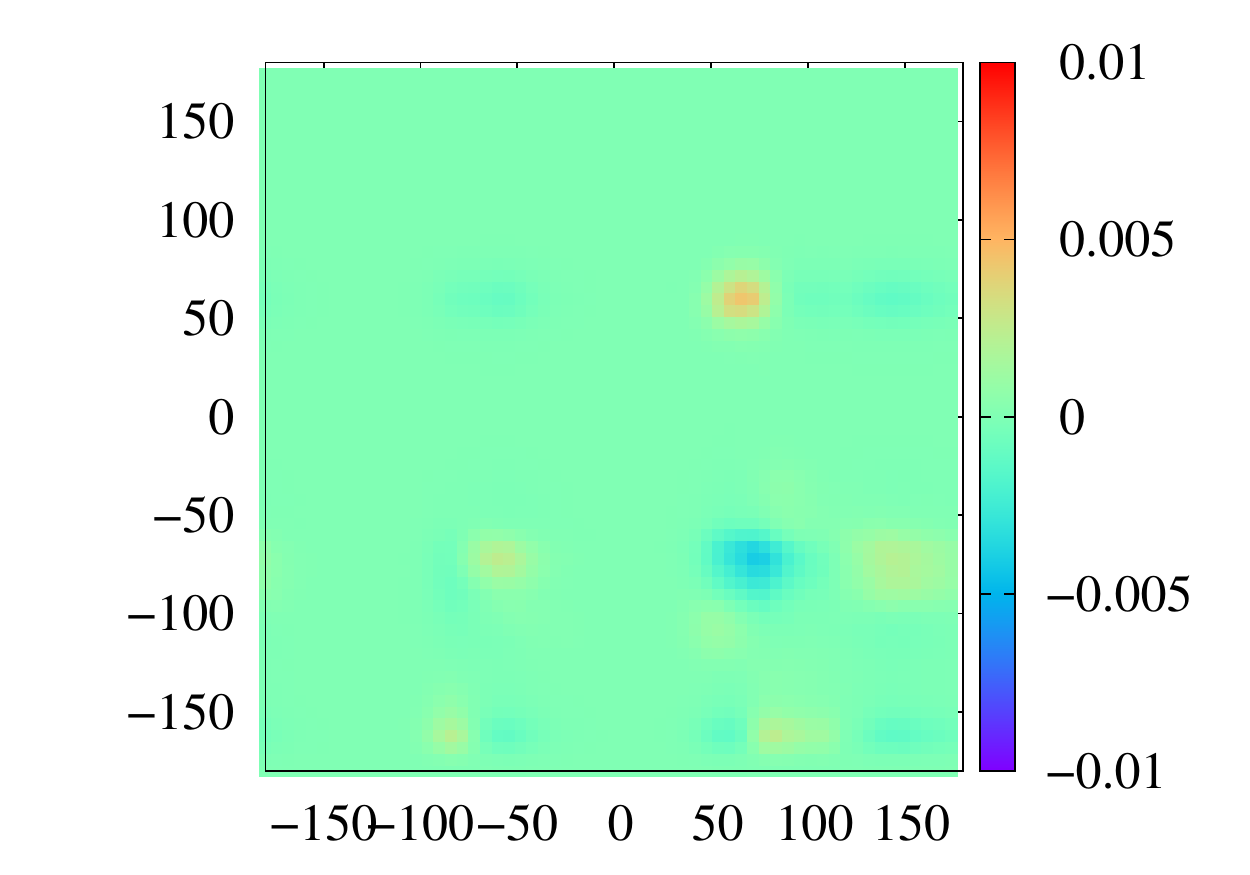}}\\
\subfloat[]{\includegraphics[width=1.7in]{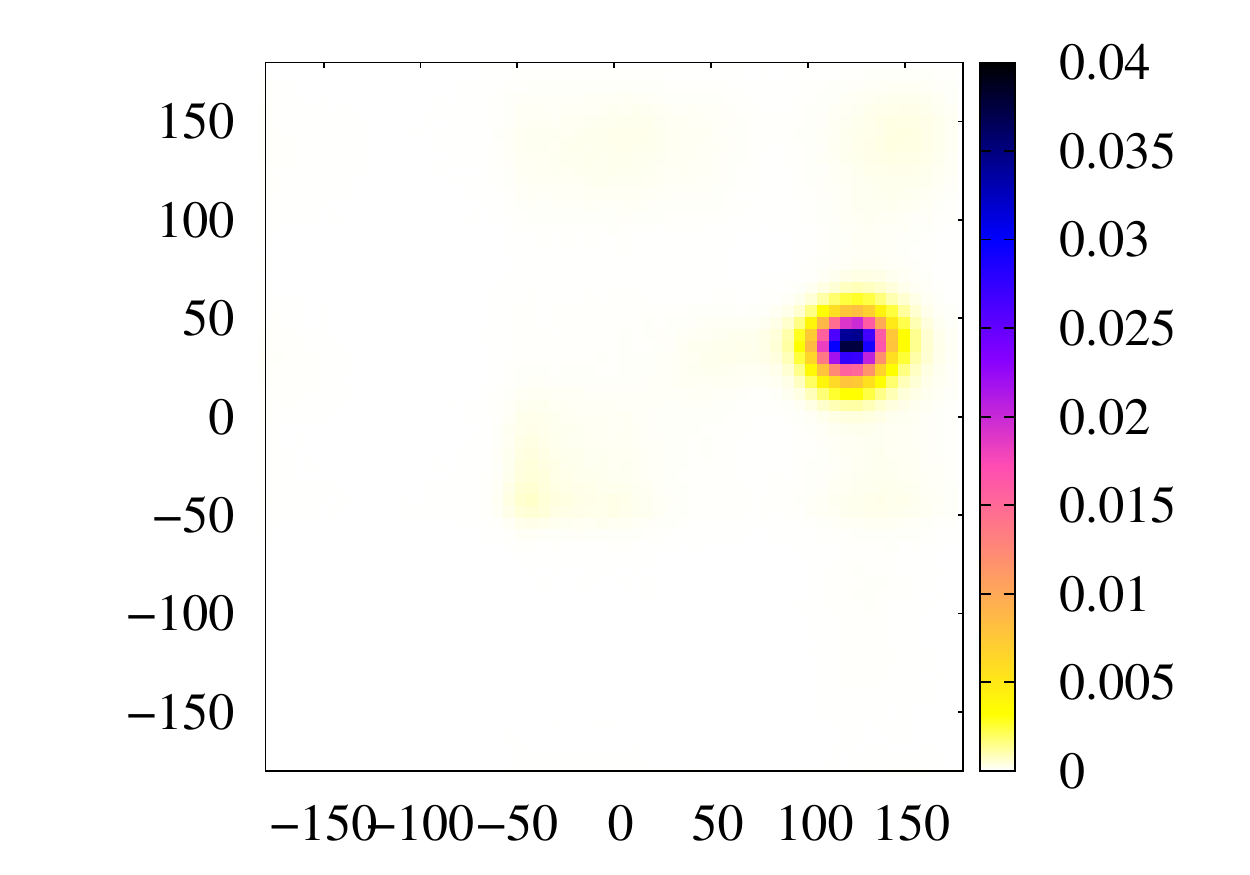}}
\subfloat[]{\includegraphics[width=1.7in]{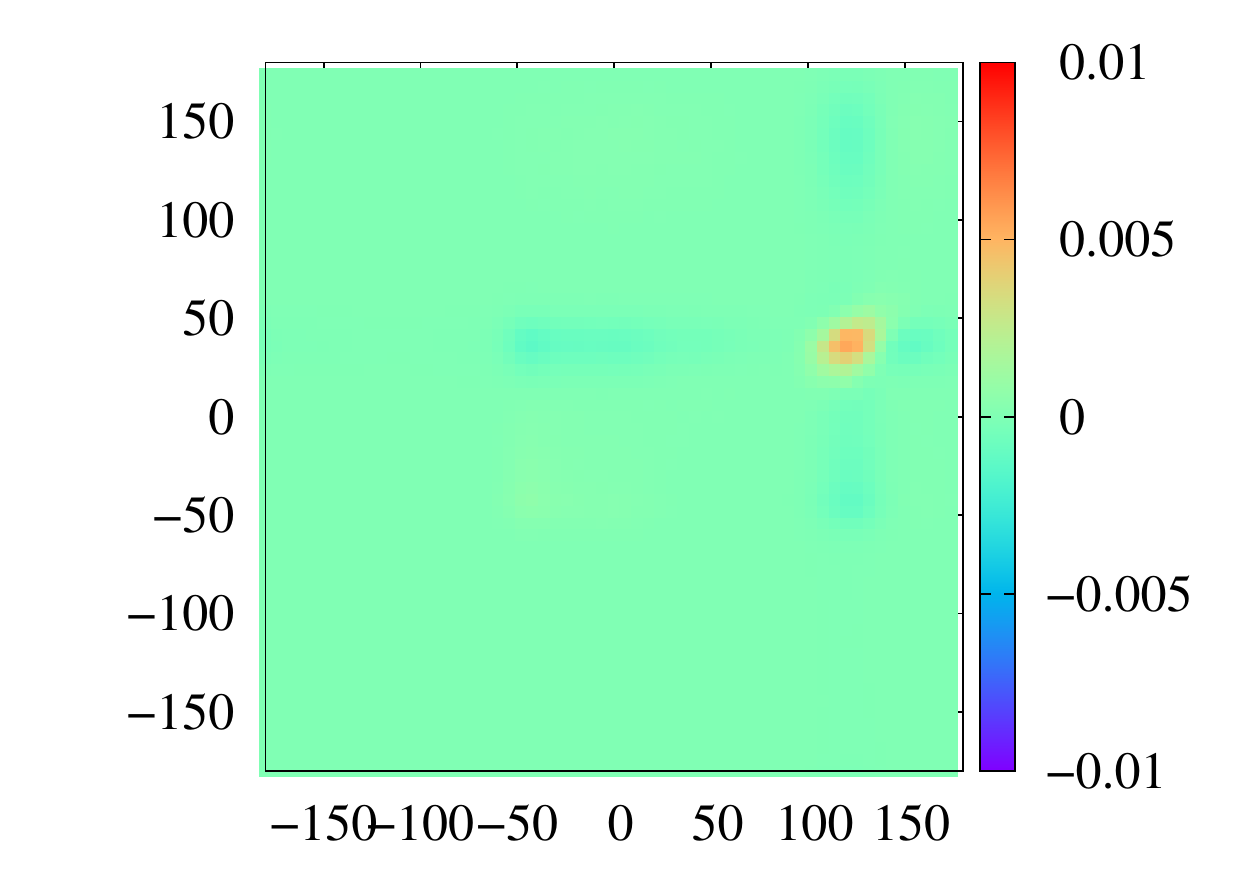}}\\
\subfloat[]{\includegraphics[width=1.7in]{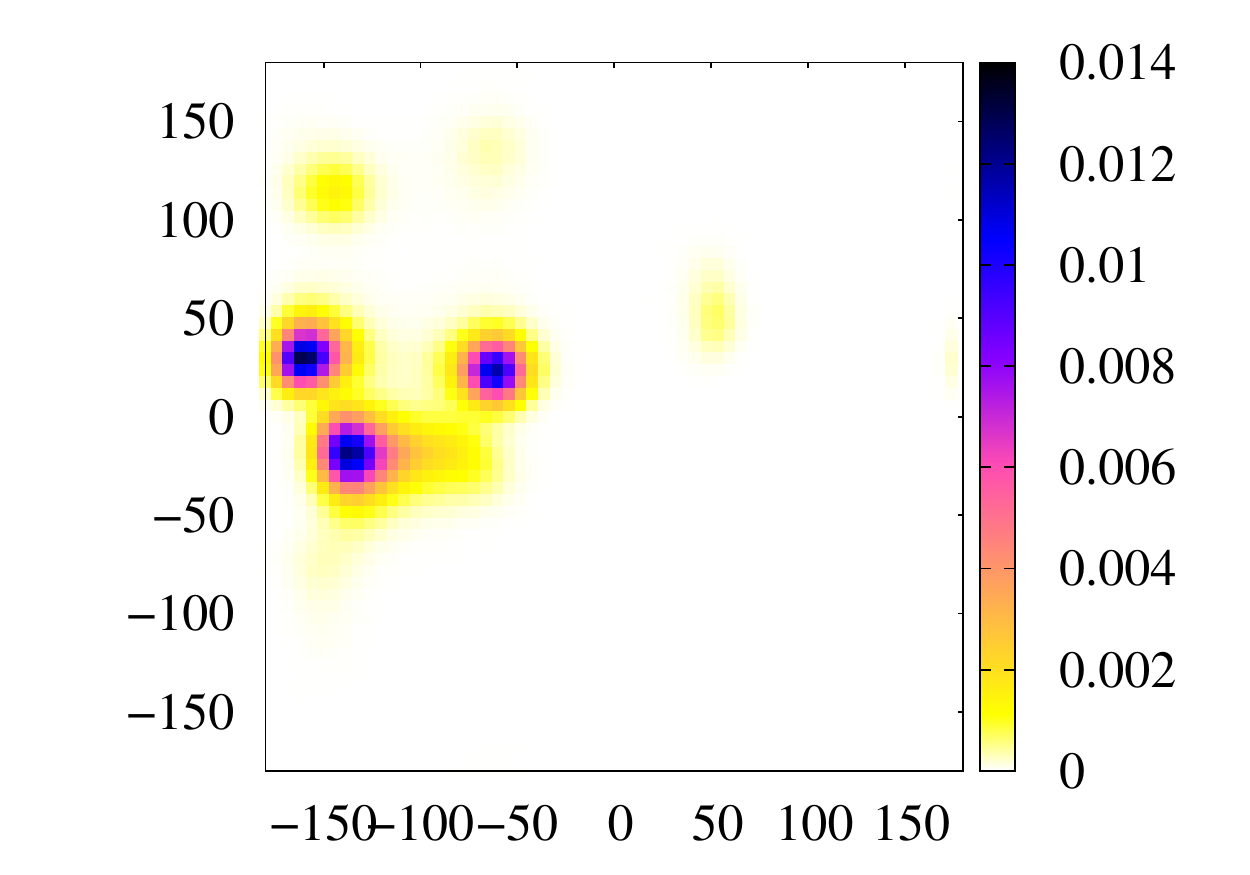}}
\subfloat[]{\includegraphics[width=1.7in]{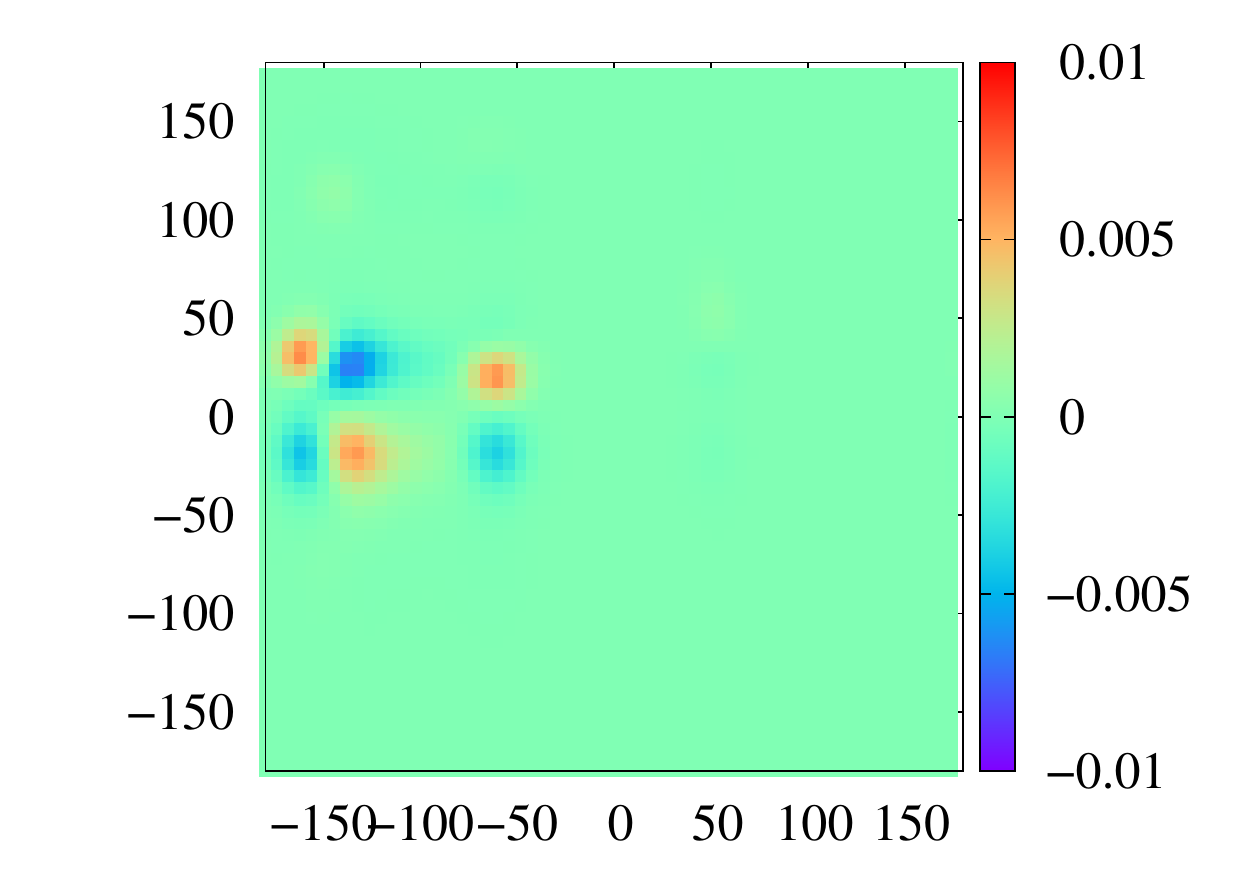}}\\
\subfloat[]{\includegraphics[width=1.7in]{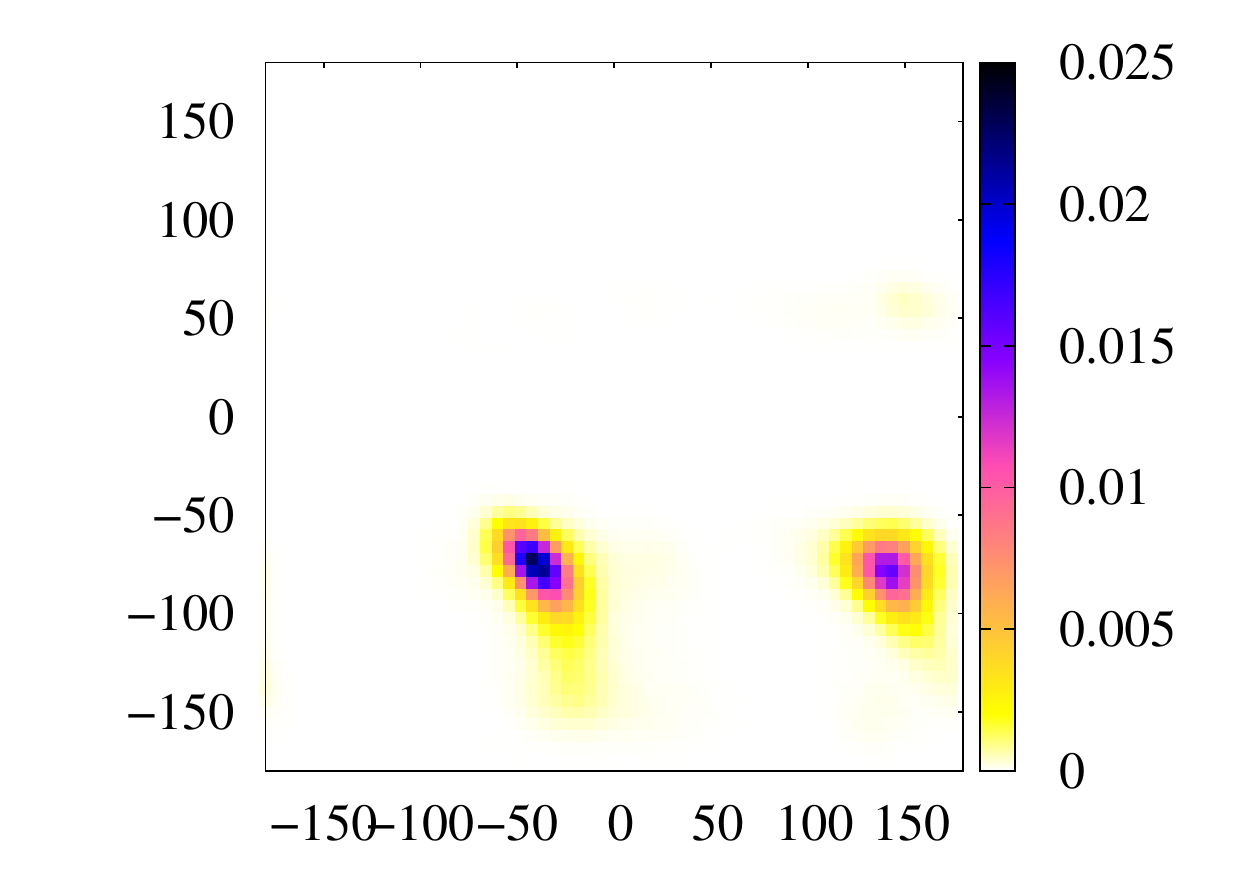}}
\subfloat[]{\includegraphics[width=1.7in]{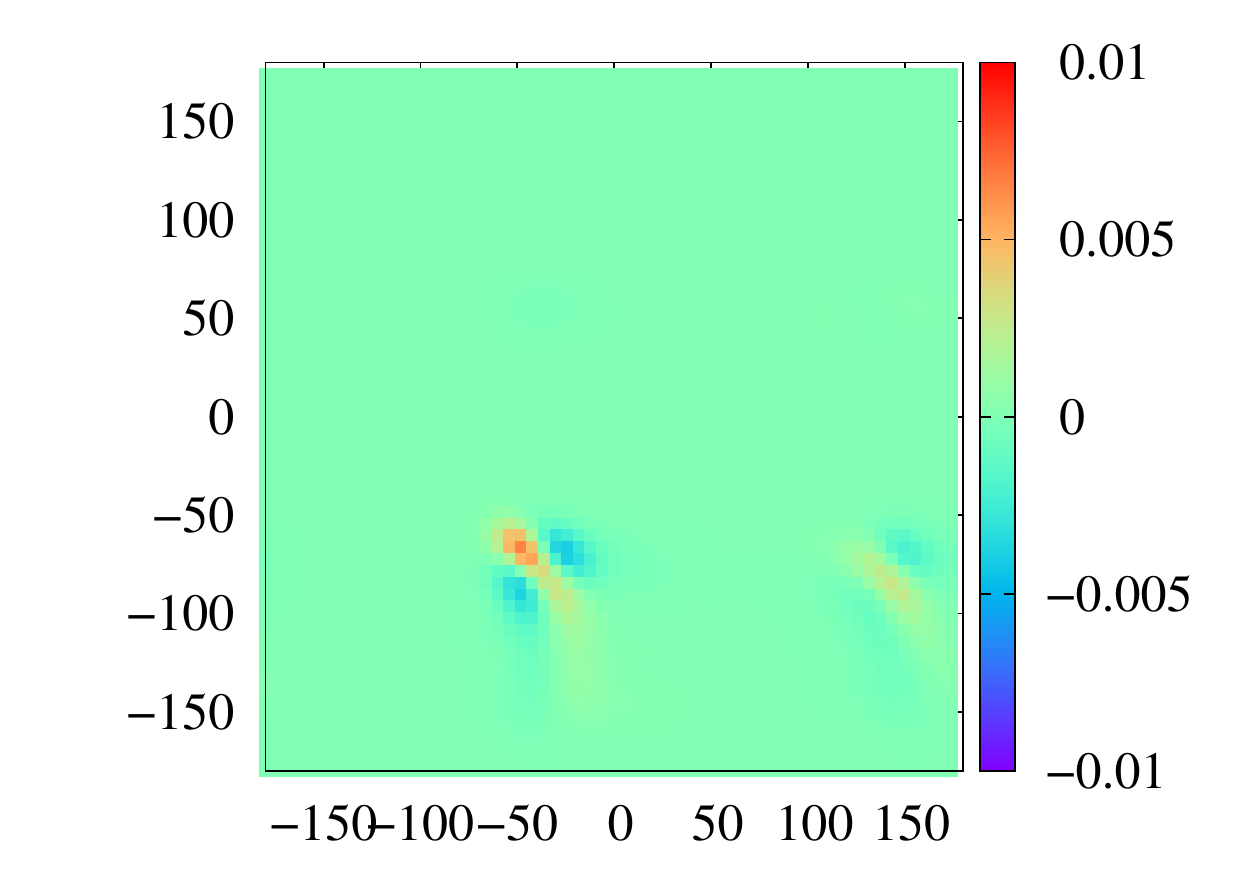}}\\
\caption{Joint distributions (the left column) and distribution differences (the right column) for five selected BTPs. a) and b) are for a typical BTP with torsional state transitions but negligible nonlinear contributions to the pair correlation. All the remaining are for four different types of heterogeneous linear correlations in BTPs with overall strong non-linear correlations} 
\label{fig:pdp}
\end{figure}

One unresolved issue is that there are some BTPs that both fall on (or locate very closely to) the contour line and have torsional state transitions in one (or both of in very few cases) the participating torsion(s). Such observations were made for full data sets of most proteins analyzed (data not shown). While the above analyses clearly demonstrated that torsional state transitions of participating torsions in a correlated BTP are necessary but not sufficient to generate non-linear correlations, we did not know for BTPs with observed torsional state transitions in participating torsions, what differentiate those BTPs fall onto (or locate closely to the contour line) from those locate far away and above from the contour line on the corresponding $MI$ vs. $r$ plot. To disentangle these complications, we selected many BTPs that locate on various positions on the $MI$ vs. $r$ plot for each protein, and examined corresponding joint distributions $p(x,y)$ and distribution difference $\Delta p(x,y) = p(x,y) - p(x)p(y)$($x$ and $y$ are selected $\phi$ or $\psi$) as shown in Fig. \ref{fig:pdp}. It was found that for BTPs locate above the contour line, linear correlations between the two participating torsions are highly heterogeneous, and in extreme cases canceling, and therefore may not be properly represented by a single linear correlation coefficient. More specifically, by heterogeneity we refer to the fact that for two variables $x$ and $y$ with respective domain of definition, their correlation varies on different subdomain of definition. While for BTPs fall onto or locate in the vicinity of the contour line, more homogeneous linear correlations were observed. This is consistent with observations in Fig. \ref{fig:subset} that for BTPs locate far from the contour line in the collective trajectory set, their positions varies greatly for different trajectory subsets. Correspondingly, homogeneity was observed for BTPs that have torsional state transitions but fall on the contour line in $MI$ vs. $r$ plot (Fig. \ref{fig:pdp}). One can imagine that for these BTPs, we will only observe linear correlations of essentially the same sign. Of course , they may be of widely different magnitudes when the linear correlation observed in the collective trajectory set is mainly caused by torsional state transitions and the participating torsions correlate weakly within each local torsional state. Based on this idea, we constructed similar $MI$ vs. $r$ plots for trajectory subsets of some selected BTPs that have torsional state transitions and fall on the contour line. Indeed, this is what we observed and BTPs remain on (or locate closely to) the contour line in all trajectory subsets (data not shown). Therefore, nonlinear contributions to BTP correlations are essentially different extent of linear correlation heterogeneity.     

\section{Discussions}
\subsection{Potential functional relevance of SLR non-linear BTP correlations and challenges}
From a potential functional point of view, proteins with diverse and significant SLR correlations may be utilized to transmit widely different signals upon different stimuli, and proteins with few SLR correlations may not be versatile in transmitting information over long distances, or at most transmit highly specific and dedicated signals.   
The biological implication is that for a protein with diverse significant SLR backbone torsional correlations executed by loop residues, plenty of opportunity exist for designing molecular agent to modulate its functions allosterically. Considering the paramount importance of flexible loop residues in coordinating and participating a wide variety allosteric interactions\cite{Ma2011,Fuxreiter2014,Motlagh2014}, and the emerging superiority of drug targeting allosteric sites\cite{Panjkovich2012,Nussinov2013,Nussinov2013a,Goncearenco2013,Pei2014,DiPaola2015}, the SLR non-linear correlations of loop residues are of far reaching potential importance in future manipulation of biological systems. However, to fully realize the potential of such versatile SLR, one need to have the capability of predicting such correlations on the one hand, and to understand the mechanism of how information transmit from one site to a distal site in a non-linear way on the other hand. Both are significant challenges that need to be addressed and are briefly discussed below. Firstly, despite the fact that with steady expected increase of computational power, sub-millisecond to milliseconds MD simulations are expected to be routine in a decade, the fact that we identified SLR nonlinear correlations does not implicate that we may accurately predict such correlations through extensive MD simulations. The major concern is the quality of force fields in describing such SLR dynamical correlations since we essentially have no reliable reference to perform corresponding optimizations. This is in contrast to the availability of highly accurate small molecule experimental data and the protein data bank\cite{} for validation of parameters describing approximate harmonic interactions (e.g. bonded-interactions) and rotameric distributions. The other possible way is to utilize machine learning technique once we have sufficient reliable data of such SLR nonlinear torsional correlations, which is unfortunately not available for the time being and the worse news is that we do not even have a well-validated general methodology to generate such data. Secondly, backbone torsions in stable secondary structures mainly exhibit harmonic intra-well dynamics and linear correlations that are on relatively short time scales (nanoseconds or shorter), while nonlinear SLR backbone torsional pair correlations are associated with aharmonic torsional state transitions that occur on much longer and widely different time scales (ranging from tens of nanoseconds up to multiple micron-seconds and beyond as observed in MD simulations). Therefore, if distal nonlinear BTP correlations were indeed transmitted through stable secondary structures, it should not be harmonic vibrational motions that contribute predominantly to linear pair correlations among on-path backbone torsions in corresponding secondary structures. Of course, SLR linear correlations are likely to be transmitted by strong linear correlations among local backbone torsional pair correlations in stable secondary structures. However, as shown in Fig. \ref{fig:dist} and Fig. \ref{fig:distss}, significant SLR backbone torsional correlations are predominantly nonlinear. Despite many insightful studies that have been carried out to achieve mechanistic and/or operational understanding of the signal transmission in allostery and to identify on-path communicating residues\cite{Bouvignies2005, Dubay2011, Selvaratnam2011,Cukier2011,Long2011,Mitternacht2011, Mitternacht2011a,England2011,Gerek2011,Panjkovich2012,Sfriso2012,Long2012,Panjkovich2014,Morra2014,McLeish2015,DiPaola2015}, this time scale issue remain to be tackled for improved understanding of how SLR non-linear correlations are mediated. 

\subsection{Time scale relevance and spurious correlations} 
\begin{figure}
\centering 
\subfloat[a]{\includegraphics[width=3.5in]{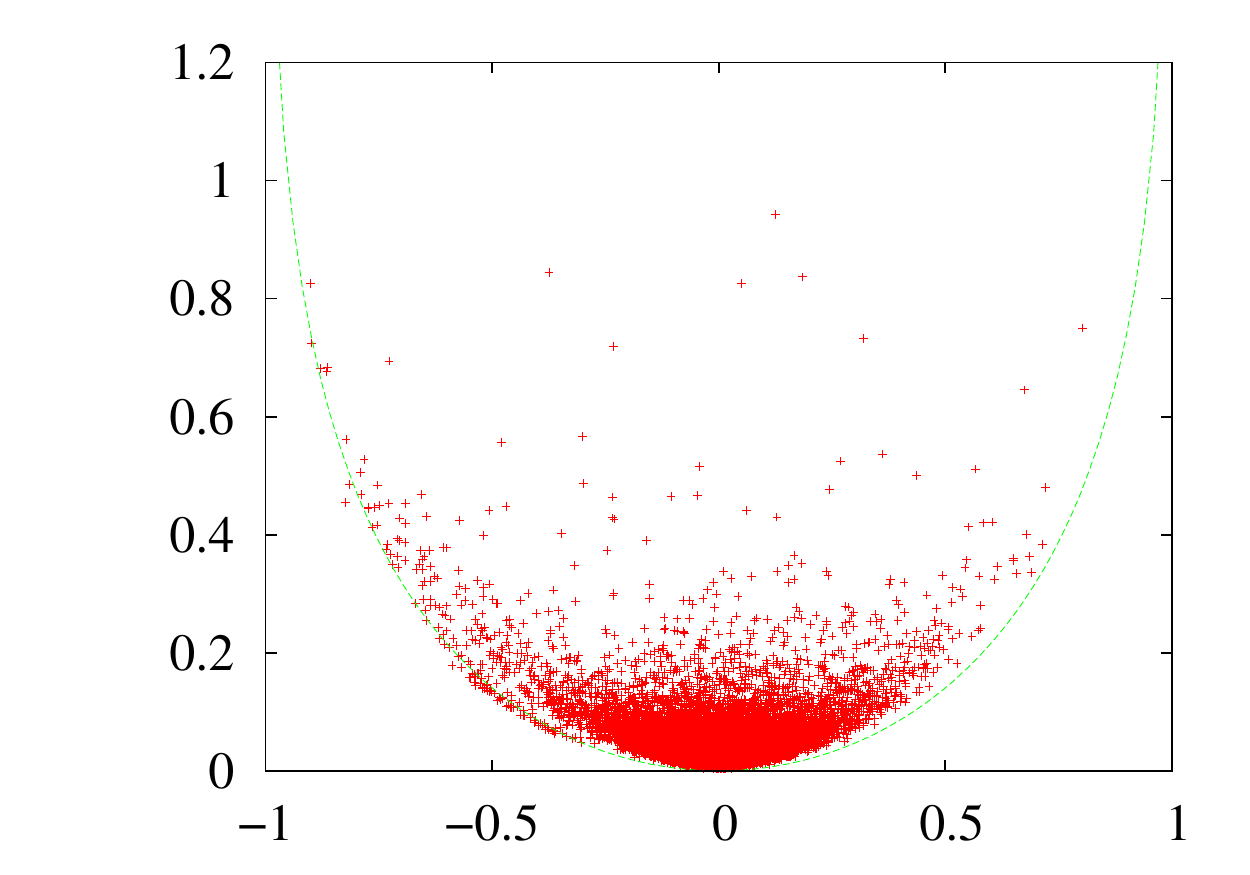}}
\subfloat[b]{\includegraphics[width=3.5in]{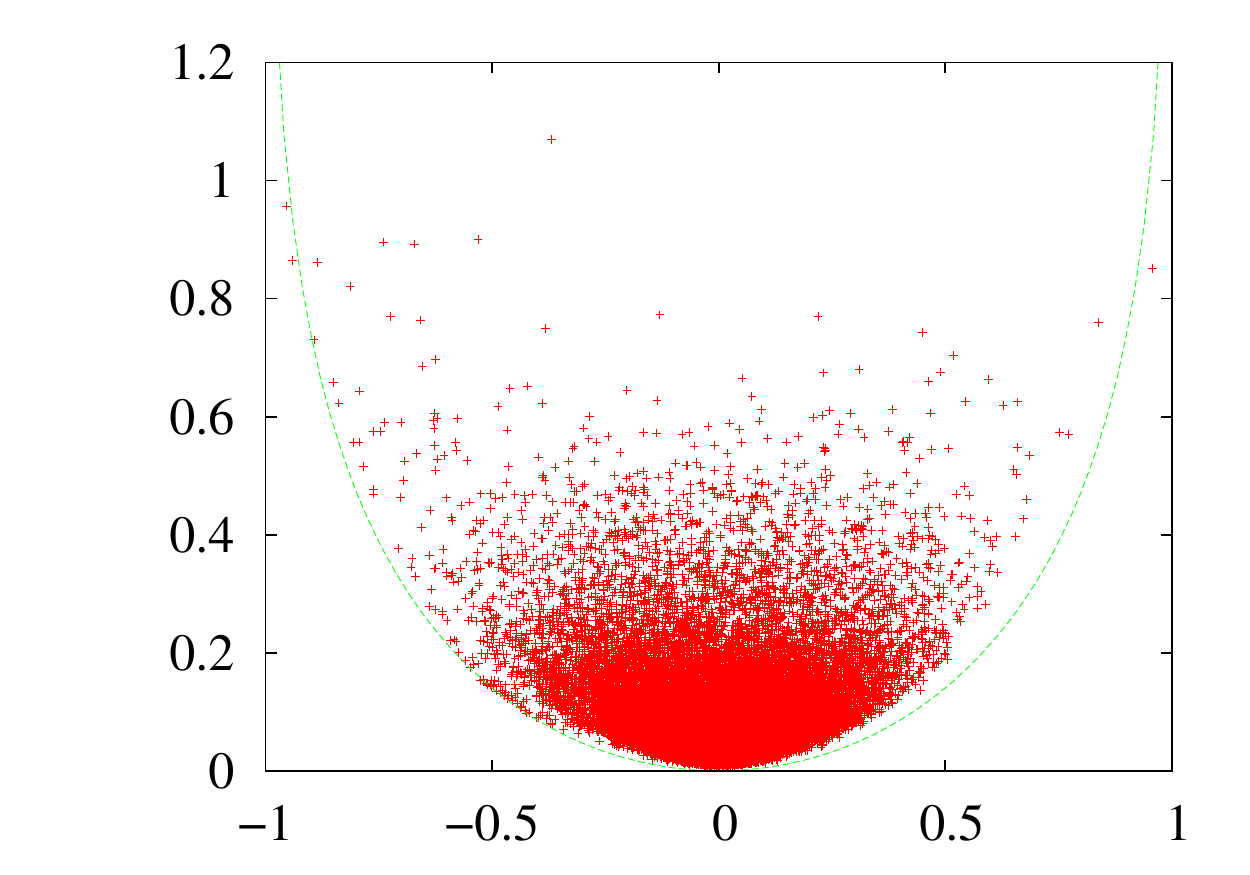}}\\
\subfloat[c]{\includegraphics[width=3.5in]{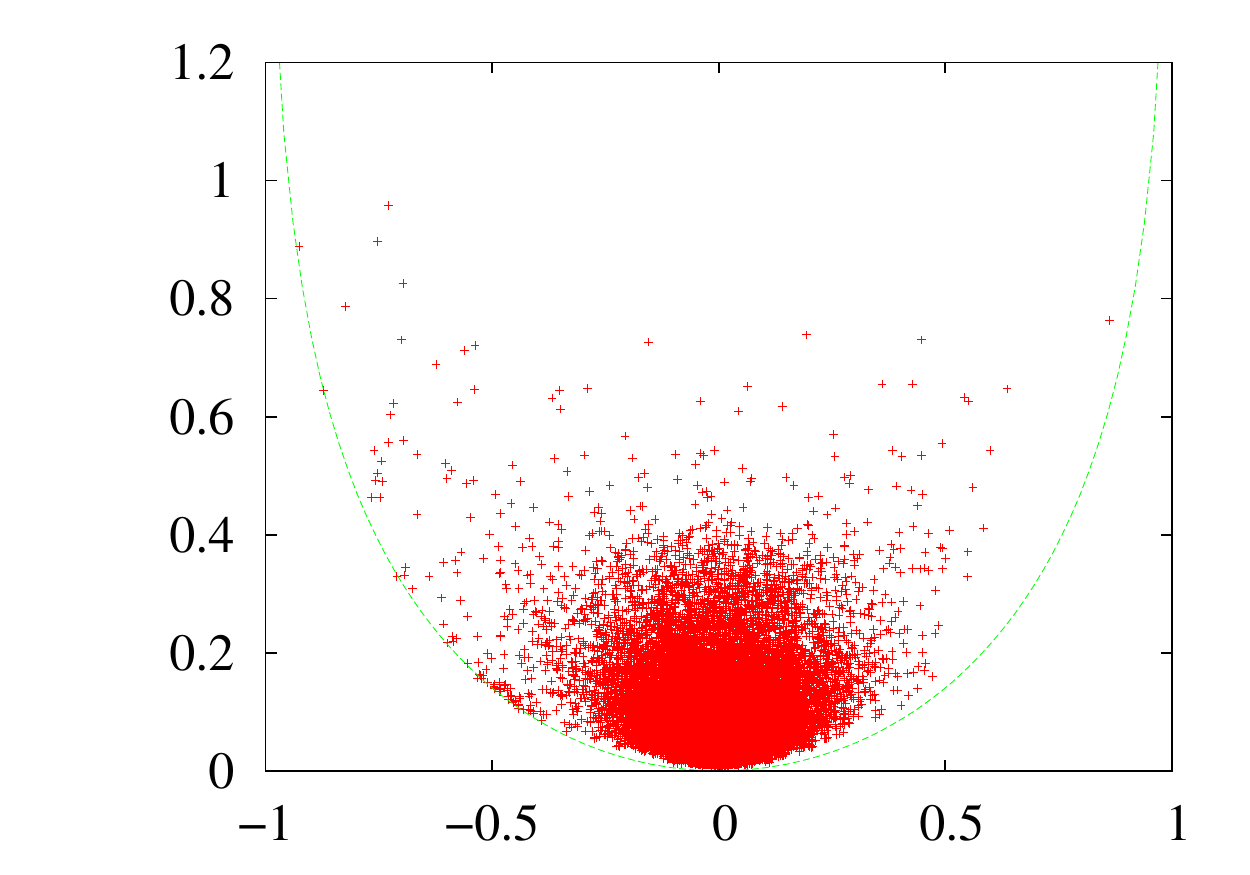}}
\subfloat[d]{\includegraphics[width=3.5in]{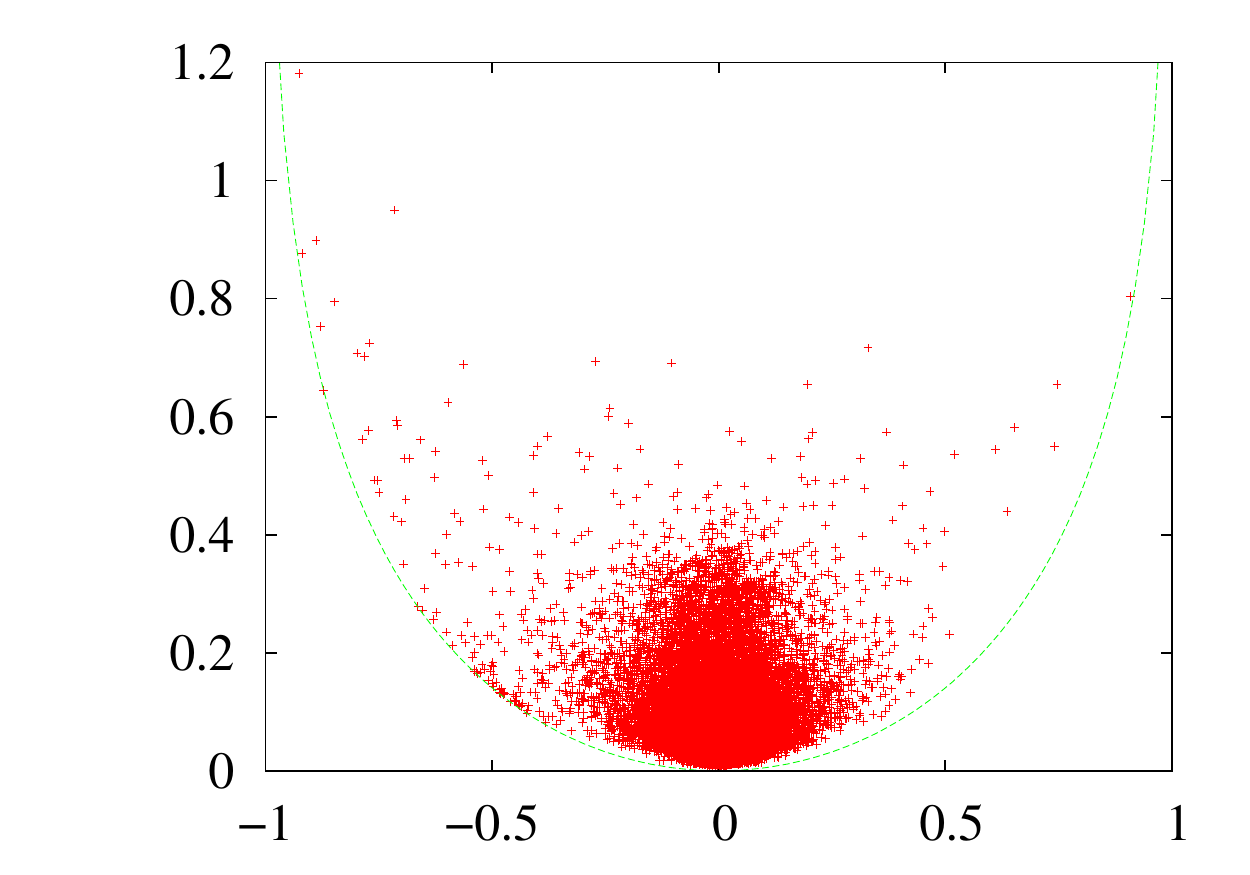}}
\caption{$MI$ vs. $r$ plots of hen egg white lysozym constructed from trajectory set of a) 1, b)10, c)100 and d)1000 independent $100$-$ns$ trajectories. } 
\label{fig:timescale}
\end{figure}

Non-linear protein BTP correlations, which is demonstrated by our analysis of MD trajectories to occur on longer time scales and larger spatial distances than linear correlations, are strong candidates for mediation of allosteric interactions. Since torsional state transitions are essential for strong SLR non-linear BTP correlations, and apparently different time scales are associated with torsional state transitions for different torsional DOFs, it is therefore important to specify time scales when one is interested in correlations of a given BTP. When small MD trajectory set (up to $\sim$10$ns$) with snapshots interval on $ps$ time scales is utilized for analysis, the results would likely to be dominated by strong linear correlations associated with harmonic local motions. Therefore, such analyses are not likely to be insightful for disclosing mechanisms of many functionally important allosteric interactions. Indeed, short time scale linear correlation based network analysis was found to be not effective\cite{}. A more latent problem associated with relatively long time scale correlations are spurious correlations, which remains a grand challenge despite many discussions before\cite{}. For illustration, we constructed $MI$ vs. $r$ plots for lysozyme based on different trajectory sets as shown in Fig. \ref{fig:timescale}. One would immediately conclude that a significant fraction of non-linear correlations observed in Fig. \ref{fig:timescale}b are spurious since they disappear in larger trajectory sets (Fig. \ref{fig:timescale}c,d). However, without larger trajectory sets, we usually have no reliable way of identifying spurious correlations from genuine ones. We may have the following thought experiment. Let's assume we have two independent MD trajectories $A$ and $B$ of the same length for two different proteins and pick a torsion $a$ in $A$ and a torsion $b$ in $B$, if torsional state transitions of $a$ and $b$ occur on very similar time scales that happen to be comparable with the total length of the two trajectories, it is likely for us to observe a strong correlation between $a$ and $b$. Such correlation is spurious since we know that there is no physical forces to coordinate torsional state transitions between $a$ and $b$, they must go out of phase gradually and eventually lose correlation if the observation was sufficiently long (or sufficiently many observations were made). However, for two DOFs that are in the same protein molecule and we do not have sufficiently long (or many) observation(s), we have no way of differentiating a genuine correlation from a spurious correlation unless we have the ability to identify physical interaction networks mediating arbitrarily given pairs of DOFs in a molecule, which is just as difficult, if not more, a task itself. We might be tempted to believe that a correlation for a BTP is genuine if we observed many torsional state transitions for the participating torsions. However, we can never be sure that there might be an much slower latent DOF, transition of which has not been observed but may destroy or strengthen correlation for our interested pair. Similarly, we might be tempted to speculate that an observed correlation for a BTP is likely to be spurious if only a few torsional state transitions were observed for participating DOFs, however, such correlations are equally likely to be genuine. Fortunately, for many interesting and important biomolecular systems, there are experimental means to estimate time scales of key molecular events, and such time scale information would be of great help in differentiating spurious correlations from genuine ones. 

It is important to note that convergence of BTP correlations (or simulation) is dependent upon our interested part of free energy landscape. For example, when our major interest is native conformational transitions, we do not need to observe folding/unfolding events that are likely to be on much longer time scales. A more naive source of spurious correlation is simply due to limited number of independent data points. With only a few data points available, it is highly likely to obtain spurious correlations. In this study, we utilized extensive MD trajectories (ranges from a few $\mu s$ to hundreds of $\mu s$) and a simple random permutation calculation suggest that our results do not suffer from trivial lack-of-data spurious correlations.
     
\subsection{Instability of circular linear correlations} 
Due to the circular property of torsion angles, brute force calculation of Pearson correlation coefficients is not possible for two torsions, and the following circular version is utilized in this study and many others\cite{}. However, when examine carefully the equation utilized for calculating the mean angle:
\begin{equation}
\bar{x}^{circular} = arctan\bigg ( \bigg[ \sum_{i=1}^Nsinx_i \bigg], \bigg[ \sum_{i=1}^Ncosx_i  \bigg] \bigg)
\end{equation}
it is apparent that when a torsion has a two peak distribution with $180\deg$ difference and approximately equal weights, summation of both sine and cosine terms essentially vanish and the calculated results only reflect local noises. We did observe such instability of circular linear correlations in our analysis
, three subsets of a BTP with essentially similar joint distribution ($p(x,y)$) and distribution difference ($\Delta p(x,y) = p(x,y) - p(x)p(y)$) have dramatically different calculated linear correlation coefficients, ranging from strongly negative linear correlation, to nearly no linear correlation and strong positive linear correlation. We repeated this problem with simulated data, and the results confirmed our speculated problem that any double peak with $180^{o}$ difference will have this problem (data not shown). Unfortunately, it seems to be difficult to come up with a new formulation that are free of this or other similar problems. Therefore, it is essential to be cautious with linear correlation obtained from circular analysis. However, mutual information is free of such instability in addition to the fact that it is linearly related to entropy. 

\section{Conclusions}
In summary, we analyzed extensive MD simulation trajectories for ten proteins of different sizes and folds, and found that significant SLR nonlinear BTP correlations exist in most of studied proteins and are predominantly executed by loop residues. In contrast, significant linear correlations are limited to shorter spatial range and time scales, and are more likely to associate with residues in stable secondary structures. Aharmonic torsional state transitions of participating torsions are essential for such nonlinear correlations, which correspondingly occur on widely different and longer time scales. It is important to note that for a correlated BTP, torsional state transitions are necessary but not sufficient for observation of significant nonlinear correlations, which is directly caused by heterogeneous linear correlations of participating torsions in different torsional states. Considering the tremendous role of loop residues in participation of biological activities and in transmission of signals, our findings implicate rich novel possibilities in modulation of biological activities. Meanwhile, time scale difference between SLR nonlinear correlations and local harmonic dynamics warrants further investigations on transmission of allosteric signals across single or multiple protein structural domains.   

\begin{acknowledgement}
This research was supported by National Natural Science Foundation of China under grant number 31270758, and by the Research fund for the doctoral program of higher education under grant number 20120061110019. 
\end{acknowledgement}

%

\bibliography{../../Correlation}

\newpage
\captionsetup[subfigure]{labelformat=empty}

\newpage

\newpage

\end{document}